\documentclass[reprint, superscriptaddress, nofootinbib, aps, twocolumn, prd]{revtex4-2}

\usepackage[english]{babel}
\usepackage{amsmath}
\usepackage{amsthm}
\usepackage{amssymb}
\usepackage{bm}
\usepackage{graphicx}
\usepackage{physics}
\usepackage[dvipsnames]{xcolor}
\usepackage[colorlinks=true, allcolors=Blue]{hyperref}
\usepackage{tensor}
\usepackage[capitalise]{cleveref}
\usepackage{acro}
\usepackage[normalem]{ulem}

\renewcommand{\Re}{\mathrm{Re}}
\DeclareMathOperator*{\argmax}{arg\,max}

\numberwithin{equation}{section}
\renewcommand\theequation{\arabic{section}.\arabic{equation}}

\DeclareAcronym{GW}{short = GW, long  = gravitational wave}
\DeclareAcronym{EM}{short = EM, long  = electromagnetic}
\DeclareAcronym{BH}{short = BH, long  = black hole}
\DeclareAcronym{GSHE}{short = GSHE, long  = gravitational spin Hall effect}
\DeclareAcronym{ODE}{short = ODE, long  = ordinary differential equation}
\DeclareAcronym{WKB}{ short = WKB, long  = Wentzel–Kramers–Brillouin}
\DeclareAcronym{GR}{short = GR, long  = general theory of relativity}
\DeclareAcronym{AGN}{short = AGN, long  = active galactic nucleus}
\DeclareAcronym{SNR}{short = SNR, long  = signal-to-noise ratio}
\DeclareAcronym{LVK}{short = LVK, long  = LIGO-Virgo-Kagra}
\DeclareAcronym{GO}{short = GO, long  = geometrical optics}
\DeclareAcronym{CE}{short = CE, long  = Cosmic Explorer}
\DeclareAcronym{ET}{short = ET, long  = Einstein Telescope}

\begin{document}

\title{Frequency- and polarization-dependent lensing of gravitational waves in strong gravitational fields}

\author{Marius A. Oancea}
\email{marius.oancea@univie.ac.at}
\affiliation{University of Vienna, Faculty of Physics, Boltzmanngasse 5, 1090 Vienna, Austria}
\author{Richard Stiskalek}
\email{richard.stiskalek@physics.ox.ac.uk}
\affiliation{Astrophysics, University of Oxford, Denys Wilkinson Building, Keble Road, Oxford, OX1 3RH, UK}
\affiliation{Universit\"{a}ts-Sternwarte, Ludwig-Maximilians-Universit\"{a}t M\"{u}nchen, Scheinerstr. 1, 81679 M\"{u}nchen, Germany}
\author{Miguel Zumalac\'arregui}
\email{miguel.zumalacarregui@aei.mpg.de}
\affiliation{Max Planck Institute for Gravitational Physics (Albert Einstein Institute), Am M\"uhlenberg 1, D-14476 Potsdam, Germany}

\begin{abstract}
The propagation of gravitational waves can be described in terms of null geodesics by using the geometrical optics approximation. However, at large but finite frequencies the propagation is affected by the spin-orbit coupling corrections to geometrical optics, known as the gravitational spin Hall effect. Consequently, gravitational waves follow slightly different frequency- and polarization-dependent trajectories, leading to dispersive and birefringent phenomena. We study the potential for detecting the gravitational spin Hall effect in hierarchical triple black hole systems, consisting of an emitting binary orbiting a more massive body acting as a gravitational lens. We calculate the difference in time of arrival with respect to the geodesic propagation and find that it follows a simple power-law dependence on frequency with a fixed exponent. We calculate the gravitational spin Hall-corrected waveform and its mismatch with respect to the original waveform. The waveform carries a measurable imprint of the strong gravitational field if the source, lens, and observer are sufficiently aligned, or for generic observers if the source is close enough to the lens. We present constraints on dispersive time delays from GWTC-3, translated from limits on Lorentz invariance violation. Finally, we address the sensitivity of current and future ground detectors to dispersive lensing. Our results demonstrate that the gravitational spin Hall effect can be detected, providing a novel probe of general relativity and the environments of compact binary systems.
\end{abstract}

\maketitle

%%%%%%%%%%%%%%%%%%%%%%%%%%%%%%%%%%%%%%%%%%%%%%%%%%%%%%%%%%%%%%%%%%%%%%%%%%%%%%%

\section{Introduction}

The first detection of \acp{GW} -- GW150914 -- by the Advanced LIGO observatory marked the beginning of the new era of gravitational-wave astronomy~\citep{TheLIGOScientific:2014jea, Abbott:2016blz}. \acp{GW} carry information about their source, but also imprints of the spacetime on which they travel. Observable sources of \acp{GW} emit over a wide range of frequencies~\citep{Flanagan2005}. As an example, the aforementioned GW150914 was detected from $\sim35$ to $250\,\mathrm{Hz}$. Its wavelength and that of any signal detectable by \ac{LVK}, remains orders of magnitude larger than that of the longest \ac{EM} signal capable of crossing the atmosphere ($\sim10\,\mathrm{MHz}$). Therefore, \acp{GW} have the potential to detect novel propagation effects at low frequencies, particularly when their wavelength approaches characteristic lengths of physical systems -- e.g. the Schwarzschild radius of a black hole or other gravitational lens. In these cases, the propagation of \acp{GW} might deviate slightly from the standard predictions of \ac{GO}~\cite{Takahashi:2003ix,Caliskan:2022hbu,Tambalo:2022wlm}.

The \ac{GO} approximation assumes that the wavelength is negligible compared to all other length scales of the system. Mathematically, this is the infinite frequency limit in which the evolution of either Maxwell or linearized gravity field equations is approximated by a set of \acp{ODE} instead of a set of partial differential equations. In this approximation, rays propagate along null geodesics, and the evolution of the field is approximated by transport equations along rays. Effects beyond the \ac{GO} approximation are well known in optics, where spin-orbit coupling\footnote{In this paper, spin-orbit coupling (see, e.g., Ref.~\cite{SOI_review}) refers to the dynamics of wave packets with internal structure, where the spin represents the internal degree of freedom of the wave packet (i.e., polarization), while the orbital part refers to the motion of the wave packet as a whole. Thus, this should not be confused with spin-orbit couplings arising in the dynamics of black hole binaries during the coalescence process \cite{Campanelli2006}.} leads to polarization-dependent propagation of \ac{EM} wave packets~\cite{OpticalMagnus, SHE-L_original, SHE_original, Bliokh2004, Bliokh2004_1, Duval2006, Hosten2008, Aiello2008, Bliokh2008, Bliokh2009}. This is known as the spin Hall effect of light~\cite{SOI_review, SHEL_review} and has been observed in several experiments~\cite{Hosten2008,Bliokh2008}. A similar effect -- the \ac{GSHE}~\cite{SHE_QM1,SHE_GW,GSHE2020,Frolov2020,GSHE_GW,Oanceathesis,Harte_2022,GSHE_Dirac} -- was predicted for wave packets propagating in curved spacetime and has been widely studied using various theoretical methods~\cite{Frolov, Frolov2, covariantSpinoptics, Harte2018, spinorSpinoptics, spinorSpinoptics2, GSHE2020, shoom2020, Frolov2020, GSHE_GW, audretsch,rudiger,SHE_QM1, SHE_Dirac, SHE_GW,souriau1974modele, saturnini1976modele, Duval,Duval2018,Duval2019} (see Refs.~\cite{GSHE_reviewCQG,GSHE_review} for a review and introduction). In this paper, we consider the \ac{GSHE} of \acp{GW} propagating on a curved background spacetime, as presented in Refs.~\cite{GSHE_GW,Harte_2022}.

The \ac{GSHE} is described by a set of effective ray equations that represent the propagation of a gravitational wave packet energy centroid up to first order in wavelength, derived as a higher-order \ac{GO} approximation using a \ac{WKB} ansatz. Within this formalism, the wave packets undergo frequency- and polarization-dependent deviations from the \ac{GO} trajectory, which can be viewed as a manifestation of the spin-orbit coupling via the Berry curvature. Moreover, the deviations are described by the same effective ray equations for both \ac{EM} and linearized gravitational fields~\citep{GSHE2020, GSHE_GW}.

\acp{GW} offer the best chance to probe the \ac{GSHE}. The \ac{GSHE} emerges as a first-order perturbation in the ratio between wavelength and the background gravitational field length scale -- the Schwarzschild radius $R_{\rm s}$. While the present day \ac{GW} terrestrial observatories have a lower limit at $\sim10\,\mathrm{Hz}$~\citep{LIGOprospects}, or equivalently wavelength of $\sim10^7\,\mathrm{m}$, radio telescopes such as the Event Horizon telescopes observe at $\sim 1.3\times10^{-3}~\mathrm{m}$~\citep{EHT}, i.e. at wavelengths orders of magnitude lower than the \ac{GW} interferometers. Therefore, there is little hope of finding observable astrophysical systems where the \ac{EM} radiation wavelength is comparable to the gravitational field length scale.

Another reason to search for \ac{GSHE} using \acp{GW} is that sources may inhabit high-curvature environments. In addition to isolated evolution of massive binary stars, \ac{GW} emitting binaries may form by dynamical encounters in a dense environment, such as a globular cluster~\cite{Gerosa:2017kvu,Rodriguez:2019huv,Fragione:2020han} or an AGN~\cite{Levin:2006uc,McKernan:2019beu,Tagawa:2019osr}. For a review of hierarchical \ac{BH} formation channels, see Ref.~\cite{Gerosa2021}. In the \ac{AGN} scenario, compact objects accumulate in the disk around a supermassive black hole~\cite{Miralda-Escude:2000kqv}. Interactions with the disk would subsequently drive them towards migration traps, stable orbits where gas torques change direction~\cite{Bellovary:2015ifg}. Migration traps could be as close as $\lesssim 10$ Schwarzschild radii of the supermassive black hole~\cite{Peng:2021vzr}. Such a ``last migration trap'' may contribute up to $\sim 1$\% of \ac{GW} events detectable by \ac{LVK}. This opens up the possibility of detecting strong field effects in \ac{GW} propagation in hierarchical triple systems, wherein the emitting \ac{BH} binary is sufficiently close to or orbiting around a massive third companion \ac{BH}.
The \ac{GSHE} may be detectable in these systems, in addition to multiple images of the merger caused by the \ac{BH} \cite{Kocsis:2012ut,DOrazio:2019fbq,Gondan:2021fpr}.

Interest in the \acp{AGN}-\ac{GW} connection boomed after LIGO-Virgo's detection of GW190521~\cite{LIGOScientific:2020iuh,LIGOScientific:2020ufj}, a binary whose primary component's mass is in the pair instability gap~\cite{Farmer:2019jed}. Such a massive \ac{BH} could not have formed from stellar evolution, pointing towards a likely dynamical origin for the binary. Furthermore, the Zwicky Transient Facility detected an \ac{EM} flare in AGN J124942.3+344929 (redshift of $0.438$), $34$ days after GW190521 and with consistent sky localization. In this tentative interpretation, the \ac{BH} binary would be in a migration trap with a semi-major axis of $\sim 350$ Schwarzschild radii of the supermassive black hole, and the delay between both events would be the time required for the \ac{EM} radiation to emerge from the accretion disk of the \ac{AGN}~\cite{Graham:2020gwr}. Although suggestive, evidence for an \ac{AGN} origin of GW190521 is far from conclusive when considering LIGO-Virgo data~\cite{Nitz:2020mga,Fishbach:2020qag,Bustillo:2021tga,Estelles:2021jnz,Olsen:2021qin} or the putative \ac{EM} counterpart~\cite{Ashton:2020kyr,Palmese:2021wcv}.

The \ac{GSHE} provides a novel test of the \ac{GW} source environments, which may help establish their \ac{AGN} formation channel. An advantage of this test is that it can be performed on individual observations. In contrast, other proposed methods require either LISA-like observatory~\citep{LISA17, LISA19} to measure the orbit of the emitting binary around the background black hole~\cite{Inayoshi:2017hgw,DOrazio:2019fbq,Toubiana:2020drf,Yu:2020dlm,Sberna:2022qbn} or population studies. The latter being based on binary properties (masses, spin, eccentricity)~\cite{Sedda:2020vwo,Zevin2021,Vajpeyi:2021qsw} or associating \ac{GW} events with detected \ac{AGN} flares~\cite{Bartos:2017ggb,Palmese:2021wcv}. Although measuring the \ac{GSHE} might be possible for only a small fraction of the \ac{GW} events originating in \ac{AGN} disks, its complementarity with other methods would yield valuable insights into \ac{BH} and \ac{GW} astrophysics.

The \ac{GSHE} arises in Einstein's \ac{GR}~\cite{GSHE_GW}, but it is also similar to effects emerging in theories beyond \ac{GR}, and thus needs to be taken into account to correctly interpret tests of gravity with \acp{GW}. A nonzero graviton mass leads to a distance and frequency-dependent propagation for all \acp{GW}~\cite{deRham:2016nuf}. Some alternative theories predict environment- and polarization-dependent \ac{GW} propagation speeds -- the \ac{GW} birefringence effect~\citep{Ezquiaga2020}. This leads to a frequency-independent time delay between the $+$ and $\times$ polarization states that may either interfere in the detector or appear as two copies of the same signal if the time delay is shorter/longer than the signal, respectively.  A related effect stems from parity-breaking terms in the effective field theory of \acp{GW}. Ref.~\cite{Wang2021} searched for frequency-dependent \ac{GW} birefringence (between left and right polarized \acp{GW}), finding that only the GW190521 observation is compatible with violation of parity. All these beyond-GR effects are related to the \ac{GSHE}, although in principle distinguishable from it. Establishing a detection of the \ac{GSHE} in the \ac{GW} data would represent yet another test of gravity and additional evidence for \ac{GR} in the strong-field regime.

We demonstrate that the \ac{GSHE} can be detected in \acs{GW} sources in a hierarchical triple system, in which a stellar-mass binary is close to a much more massive companion, such as in an \ac{AGN}. The main observable signature of the \ac{GSHE} is time delay between the high- and low-frequency components of the waveform, with a correction proportional to $\sim 1/f^2$ relative to geodesic propagation. Therefore, the \ac{GSHE} may appear as an inconsistency between the higher and lower frequency parts of the waveform (e.g., in inspiral-merger-ringdown tests of \ac{GR}). A subdominant \ac{GSHE} signature is a frequency-dependent birefringence effect -- a time delay between the left- and right-polarized components. \ac{GSHE}-birefringence is further suppressed ($\sim 1/f^3$) and is likely too small to be detectable, except in fine-tuned configurations. A third signature of this scenario is the likely presence of multiple signals due to strong-field lensing by the massive \ac{BH}. The relative magnification, time delay, and sign of the \ac{GSHE} correction between these signals should allow for further means to probe the system configuration.

The paper is organized as follows. We begin by describing the \ac{GSHE} and the numerical calculation of the time of arrival delay in~\Cref{sec:methodology}. In~\Cref{sec:results}, we describe the dependence of the time of arrival delay on frequency, polarization state, and the mutual position of the source and the observer. In~\Cref{sec:results}, we demonstrate the effect of the \ac{GSHE} on a \ac{GW} waveform and its distinguishability from an uncorrected waveform. Lastly, we discuss our findings in~\cref{sec:discussion} and conclude in~\Cref{sec:conclusion}. Our results are also presented in a more compact form in the companion Letter, Ref.~\cite{GSHE_lensing_letter}.

We note that $\log$ refers to a logarithm of base $10$, $x\cdot y = x^\mu y_\mu$ denotes the inner product of $4$-vectors and, unless explicitly discussing dimension-full quantities, we set the speed of light, the gravitational constant and the Kerr \ac{BH} mass $M$ to unity, $c=G=M=1$.

%%%%%%%%%%%%%%%%%%%%%%%%%%%%%%%%%%%%%%%%%%%%%%%%%%%%%%%%%%%%%%%%%%%%%%%%%%%%%%%%

\section{Methodology}\label{sec:methodology}

We assume the existence of a \ac{GW} emitter -- a binary \ac{BH} merger -- in the vicinity of a Kerr \ac{BH}, with \ac{GW} ray trajectories passing through the strong-field regime of the background Kerr metric. We then calculate the observer time of arrival of the \ac{GSHE} trajectories, which depends on frequency and polarization, and compare it to the geodesic time of arrival. In other words, the observer detects that the waveform modes have a frequency- and polarization-dependent time of arrival that deforms the resulting waveform.

We start by reviewing the Kerr metric and \ac{GSHE} equations in~\Cref{sec:GSHEeqs}. We then present our geometric setup in~\Cref{sec:configuration} and numerical integration in~\Cref{sec:integration}. Finally, we characterize the \ac{GSHE} time delay quantities in~\Cref{sec:quant_delay} and discuss our waveform model in~\Cref{sec:waveform_modelling}.

%%%%%%%%%%%%%%%%%%%%%%%%%%%%%%%%%%%%%%%%%%%%%%%%%%%%%%%%%%%%%%%%%%%%%%%%%%%%%%%%

\subsection{Gravitational spin Hall equations}\label{sec:GSHEeqs}

We consider the background spacetime of a Kerr black hole with mass $M=1$ and spin parameter $a$, described using Boyer-Lindquist coordinates $(t, r, \theta, \phi)$~\citep[p.195]{griffiths2009}. The line element is
\begin{equation} \label{eq:metric}
\begin{split}
    \dd s^2
    =
    &- \frac{\Delta}{\Sigma} \left( \dd t - a \sin^2 \theta \dd\phi \right)^2 + \frac{\Sigma}{\Delta} \dd r^2 + \Sigma \dd \theta^2 \\
    &+ \frac{\sin^2\theta}{\Sigma} \left[ a \dd t - (r^2+a^2) \dd \phi \right]^2,
\end{split}
\end{equation}
where
\begin{subequations}
\begin{align}
    \Delta &= r^2 - 2 M r + a^2,\\
    \Sigma &= r^2 + a^2 \cos^2 \theta.
\end{align}
\end{subequations}
We also consider an orthonormal tetrad
\begin{subequations}\label{eq:kerr_orthonormal_tetrad}
\begin{align}
    e_0 &= \frac{1}{\sqrt{\Delta \Sigma} } \left[ (r^2 + a^2) \partial_t + a \partial_\phi \right], \\
    e_1 &= \sqrt{\frac{\Delta}{\Sigma}} \partial_r, \\
    e_2 &= \frac{1}{\sqrt{\Sigma}} \partial_\theta, \\
    e_3 &= \frac{1}{\sin \theta \sqrt{\Sigma} } \left( a \sin^2 \theta \partial_t + \partial_\phi \right),
\end{align}
\end{subequations}
that satisfies $(e_a)^\mu (e_b)_\mu = \tensor{\eta}{_a _b}$, where $\tensor{\eta}{_a _b}$ is the Minkowski metric. The vectors $e_a$ will be used in the definition of the \ac{GSHE} and for the prescription of initial conditions.

On the Kerr background spacetime, we consider \acp{GW} represented by small metric perturbations and described by the linearized Einstein field equations. High-frequency \acp{GW} can be described using the \ac{GO} approximation~\cite[Sec. 35.13]{MTW}, in which case their propagation is determined by the null geodesics of the background spacetime. However, at high but finite frequencies, higher-order corrections to the \ac{GO} approximation become important.

In this paper, we consider first order in wavelength corrections to the \ac{GO} approximation, wherein the propagation of \acp{GW} is frequency- and polarization-dependent. This is known as the \ac{GSHE}~\citep{GSHE_GW}, and the propagation of circularly polarized gravitational wave packets is described by the \ac{GSHE} equations~\citep{GSHE_GW, Harte_2022}
\begin{subequations}\label{eq:GSHE_ODEs}
    \begin{align}
    \dot{x}^\mu &= p^\mu + \frac{1}{p \cdot T} S^{\mu \beta} p^\nu \nabla_\nu T_\beta,
    \\
    \dot{x}^\nu \nabla_\nu p_\mu  &= -  \frac{1}{2} R_{\mu \nu \alpha \beta}  p^\nu S^{\alpha \beta},
    \end{align}
\end{subequations}
where $x^\mu(\tau)$ is the worldline of the energy centroid of the wave packet, $p_\mu(\tau)$ is the average momentum of the wave packet, the spin tensor $S^{\alpha \beta}$ describes the angular momentum carried by the wave packet and $T^\alpha$ is a timelike vector field with respect to which the energy centroid of the wave packet is defined. We eliminate the \ac{ODE} for $p_0$ by enforcing the null momentum condition $p \cdot p = 0$ along the worldline. For the circularly polarized wave packets that we consider here, the spin tensor is uniquely fixed as
\begin{equation}
    S^{\alpha \beta} = \frac{\epsilon s}{p \cdot T} \varepsilon^{\alpha \beta \gamma \lambda} p_\gamma T_\lambda,
\end{equation}
where $s = \pm 2$, depending on the state of circular polarization. In the context of the high-frequency analysis \cite{GSHE_GW, Harte_2022} leading to the above equations, the wave frequency $f$ measured by an observer with $4$-velocity $T^\alpha$ is defined as $p \cdot T = - \epsilon f$. 

The small dimensionless parameter $\epsilon$ has the same meaning as in standard high-frequency approximations in general relativity (see, for example, Refs. \cite[Sec. 1.5]{Maggiore_GWbook} and \cite[Sec. 3.2]{gravitational_lenses_book}), and is meant to keep track of the order of different terms in these expansions. In particular, the GSHE equations were derived in Ref. \citep{GSHE_GW} under the assumption that the wavelength $\lambda$ is much smaller than the length scale $L$ over which the spacetime varies significantly. Thus, the small expansion parameter is defined as $\epsilon = \mathcal{O}(\lambda / L)$, and the GSHE equations provide a reasonable approximation only when $\epsilon < 1$. In the case we are considering here, the lengthscale over which spacetime varies significantly is set by the size of the black hole, so we can take
\begin{equation}\label{eq:epsilon}
	\epsilon = \frac{\lambda}{M} = 2\frac{\lambda}{R_{\rm s}},
\end{equation}
where $\lambda$ is the wavelength of the wave packet in the rest frame of the source. This can also be expressed in dimension-full quantities as
\begin{equation}\label{eq:epsilon_SI}
	\epsilon
	=
	\frac{c^3}{G} \frac{1}{f M}
	\approx
	0.1\left(\frac{40~\mathrm{Hz}}{f}\right)\left(\frac{5\times10^4 M_\odot}{M}\right).
\end{equation}

The \ac{GSHE} equations in~\cref{eq:GSHE_ODEs} depend on the choice of a timelike vector field $T^\alpha$. The role of this vector field has been discussed in detail in Ref.~\cite{Harte_2022}, where it has been shown to have physical meaning only at the point of emission and the point of observation of a polarized ray. At these points, $T^\alpha$ can be identified with the $4$-velocity of the source and observer, respectively, and is responsible for the relativistic Hall effect~\cite{Relativistic_Hall, Stone2015}. Nevertheless, one has to choose a smooth vector field $T^\alpha$ defined everywhere in the region where the \ac{GSHE} equations are to be integrated. We discuss our choice of $T^\alpha$ in the following subsection.

%%%%%%%%%%%%%%%%%%%%%%%%%%%%%%%%%%%%%%%%%%%%%%%%%%%%%%%%%%%%%%%%%%%%%%%%

\subsection{Spatial configuration}\label{sec:configuration}

We consider a static source of \acp{GW} close to the \ac{BH} at $\bm{x}_{\rm src} = (r_{\rm src}, \theta_{\rm src}, \phi_{\rm src}) $ with a $4$-velocity $T^\alpha_{\rm src}$ and a static observer far from the BH at $\bm{x}_{\rm obs} = (r_{\rm obs}, \theta_{\rm obs}, \phi_{\rm obs})$ with a $4$-velocity $T^\alpha_{\rm obs}$. The timelike vector field $T^\alpha$ appearing in the GSHE equations \eqref{eq:GSHE_ODEs} is chosen such that
\begin{equation}
    \left. T^\alpha \right|_{\bm{x}_{\rm src}} = T^\alpha_{\rm src}
    \quad\mathrm{and}\quad
    \left. T^\alpha \right|_{\bm{x}_{\rm obs}} = T^\alpha_{\rm obs}.
\end{equation}
We start with the orthonormal tetrad $e_a$ from~\cref{eq:kerr_orthonormal_tetrad} and perform a spacetime-dependent local Lorentz boost of the orthonormal tetrad such that $(e_0)^\alpha$ maps to $T^\alpha_{\rm src}$ and $T^\alpha_{\rm obs}$ at $\bm{x}_{\rm src}$ and $\bm{x}_{\rm obs}$, respectively. We can express the boosted orthonormal tetrad $\tilde{e}_a$ as
\begin{subequations}\label{eq:boosted_tetrad}
\begin{align}
    \tilde{e}_0 &= \frac{e_0 + v e_3}{\sqrt{1 - v^2}},\\
    \tilde{e}_1 &= e_1,\\
    \tilde{e}_2 &= e_2,\\
    \tilde{e}_3 &= \frac{e_3 + v e_0}{\sqrt{1 - v^2}},
\end{align}
\end{subequations}
where
\begin{equation}\label{eq:tetrad_boost}
    v(r)
    = -\frac{a \sin{\theta_{\rm obs}}}{\sqrt{\Delta(r_{\rm obs})}} e^{- (r - r_{\rm obs})^2}
    - \frac{a \sin{\theta_{\rm src}}}{\sqrt{\Delta(r_{\rm src})}} e^{- (r - r_{\rm src})^2}.
\end{equation}
The exponential factor ensures a smooth transition between $T^{\alpha}_{\rm src}$, $\tensor{e}{_0 ^\alpha}$ and $T^{\alpha}_{\rm obs}$. We identify the timelike observer vector field in the \ac{GSHE} equations \eqref{eq:GSHE_ODEs} as $T^\alpha = (\tilde{e}_0)^\alpha$ and further justify the Lorentz boost in~\cref{sec:tetrad}.

For simplicity, we consider a static isotropic emitter of \acp{GW} in the vicinity of a massive ``lensing'' \ac{BH} that sources the background Kerr metric and a far static observer measuring the waveform (wave packet). The caveat of isotropic emission is relevant as the emission direction of the $\epsilon s$-parameterized bundle trajectories must be rotated with respect to the geodesic, $\epsilon = 0$, emission direction by an angle $\sim \epsilon s$. In this work, we do not account for the directional dependence as it is a subdominant effect. Including it would necessitate accounting for it while generating the waveform frequency modes. The Boyer-Lindquist coordinate time $t$ can be related to the static observer's proper time $\tau$ as
\begin{equation}\label{eq:coordinate_to_statobs_propertime}
    \tau = t \sqrt{- \left.\tensor{g}{_0 _0}\right|_{\bm{x}_{\rm obs}}},
\end{equation}
which we derive in~\Cref{sec:observer_proper_time}, with $\tensor{g}{_\mu _\nu}$ being the Kerr metric tensor. Throughout the rest of the paper, we denote the coordinate time as $t$ and the static observer proper time as $\tau$.

A signal with initial momentum $p_{\rm init}$ emitted by the static source with $4$-velocity $T_{\rm src}$ has a frequency $f_{\rm src}$ in the source frame. On the other hand, the static observer with $4$-velocity $T_{\rm obs}$ will measure the signal frequency as $f_{\rm obs}$. The observer, therefore, measures the signal redshifted by
\begin{equation}\label{eq:gravredshift}
    \frac{\lambda_{\rm obs}}{\lambda_{\rm src}}
    = \frac{f_{\rm src}}{f_{\rm obs}}
    = \frac{T_{\rm src} \cdot p_{\rm init}}{ T_{\rm obs} \cdot p_{\rm f}},
\end{equation}
where $p_{\rm f}$ is the wave packet's momentum when it reaches the observer. This is the common expression for gravitational redshift, which is satisfied up to first order in $\epsilon$. The $\epsilon$ dependence of the gravitational redshift originates from $p_{\rm f}$ and $p_{\rm init}$, as the initial conditions of a trajectory between two fixed spatial locations depend on $\epsilon$. We will find the $\epsilon$ dependence of the gravitational redshift to be negligible. Therefore, since this produces a uniform frequency offset and no new effect, we do not consider this further. Moreover, upon emission, the following relation is enforced,
\begin{equation}\label{eq:freq_epsilon_conservation}
    T_{\rm src} \cdot p_{\rm init} = - f_{\rm src} \epsilon = -1,
\end{equation}
where the last equality follows from~\cref{eq:epsilon}. An analogue of this condition is then satisfied along the trajectory, as discussed in Ref.~\cite{Harte_2022}.

We parameterize $p_{\rm init}$ by a unit three-dimensional Cartesian vector $\bm{k}$ expressed in spherical coordinates where $0 \leq \psi \leq \pi$ and $0 \leq \rho < 2\pi$ are the polar and azimuthal angle, respectively. The angles $\psi$ and $\rho$ represent the emission direction on the source celestial sphere, and we have that
\begin{equation}\label{eq:initial_momentum}
    p_{\rm init}
    =
    \tilde{e}_0 + \sin\psi\cos\rho \tilde{e}_1 + \sin\psi\sin\rho \tilde{e}_2 + \cos\psi \tilde{e}_3,
\end{equation}
which satisfies both~\cref{eq:freq_epsilon_conservation} and the null momentum condition $p \cdot p = 0$. The initial momentum pointing towards the \ac{BH}, i.e. with an initial negative radial component, may equally be parameterized with $k_2$ and $k_3$,
\begin{equation}\label{eq:shadow_initial_momentum}
    p_{\rm init}
    =
    \tilde{e}_0 - \sqrt{1 - k_2^2 - k_3^2}~\tilde{e}_1 + k_2 \tilde{e}_2 + k_3 \tilde{e}_3,
\end{equation}
which can be related to $\psi$ and $\rho$ as
\begin{subequations}
\begin{align}
    k_2 &= \sin \psi \sin \rho,\\
    k_3 & = \cos \psi.
\end{align}
\end{subequations}

We calculate the magnification factor $\mu$ defined as the ratio of the source area to the image area~\cite{gravitational_lenses_book,petters2012singularity, dodelson_2017}. In our case, a trajectory defines a mapping from the celestial sphere of source to the far sphere of radius $r = r_{\rm obs}$ centered at the origin, which we can write as $(\psi, \rho) \mapsto (\theta, \phi)$. The magnification $\mu$ is
\begin{equation}
    \mu
    =
    \frac{\sin \psi \dd \psi \dd \rho}{\sin \theta \dd \theta \dd \phi}
    =
    \frac{\sin \psi}{\sin \theta} \frac{1}{\det \mathbf{J}},
\end{equation}
where the Jacobian $\mathbf{J}$ is defined as
\begin{equation}\label{eq:trajectoryjacobian}
    \mathbf{J}
    =
    \pdv{(\theta, \phi)}{(\psi, \rho)}.
\end{equation}
The magnification scales a signal propagated along a trajectory by a factor of $\sqrt{|\mu|}$ and the signal parity is given as the sign of $\mu$, or equivalently the sign of $\det \mathbf{J}$. Therefore, as a consequence of the \ac{GSHE} the magnification is dependent on frequency and polarization. We will explicitly denote this dependence as $\mu(f, s)$ and the \ac{GO} magnification as $\mu_{\rm GO}$.

%%%%%%%%%%%%%%%%%%%%%%%%%%%%%%%%%%%%%%%%%%%%%%%%%%%%%%%%%%%%%%%%%%%%%%%%

\subsection{Numerical integration}\label{sec:integration}

Given a fixed source and observer, our objective is to find the connecting GSHE trajectories of the $\epsilon s$ bundle. We numerically integrate the \ac{GSHE} \acp{ODE} of~\cref{eq:GSHE_ODEs}, or the null geodesic \acp{ODE} recovered by substituting $\epsilon \rightarrow 0$, starting at coordinate time $t=0$, source position $\bm{x}_{\rm src}$ and initial wave packet momentum $p_{\rm init}(\bm{k})$ as discussed in~\Cref{sec:configuration}.

The Boyer-Lindquist coordinates contain coordinate singularities at the \ac{BH} horizon and the coordinate north and south poles. Therefore, we include the following premature integration termination conditions. First, the integration is terminated if a trajectory penetrates or passes sufficiently close by the \ac{BH} horizon, so that its radial component satisfies
\begin{equation}\label{eq:BH_horizon}
    r \leq \Delta\mathcal{H} \left(1 + \sqrt{1 - a^2}\right),
\end{equation}
where we set $\Delta\mathcal{H} = 1 + 10^{-4}$. Second, we terminate trajectories whose polar angle does not satisfy $\theta_{\rm tol} \leq \theta \leq \pi - \theta_{\rm tol}$, where $\theta_{\rm tol} = 10^{-5}$. Lastly, we optionally support early termination if the absolute value of the difference between the current and initial azimuthal angle $\Delta \phi = |\phi - \phi_{\rm src}|$ satisfies $\Delta \phi > \max(2\pi - |\phi_{\textrm{obs}} - \phi_{\textrm{src}}|, |\phi_{\textrm{obs}} - \phi_{\textrm{src}}|)$ since such solutions correspond to ones that complete more than one complete azimuthal loop around the \ac{BH}. We refer to trajectories that do not completely loop around the \ac{BH} as ``direct''.

If no early termination condition is met, we terminate the integration when the trajectory reaches the observer's radius $r_{\rm obs}$. The integrator then outputs $x_{\rm f}$ and $p_{\rm f}$, the location and momentum vectors of the trajectory at that instant. Typically, for each source-observer configuration, there exist at least two bundles that directly connect the source and the observer, with additional bundles completely looping around the \ac{BH}. Rays that loop multiple times around the \ac{BH} are a general signature of strong gravitational fields and images formed under such conditions are also referred to as retrolensing or glory effect \cite{1979A&A....75..228L,Holz_2002,bozza2010,Eiroa2012,PhysRevD.104.103011}.

We quantify whether a choice of initial direction $\bm{k}$ (and thus initial momenta) leads to a trajectory intersecting with the observer by calculating the angular distance $\Delta \sigma$ between the observer and the integrated trajectory
\begin{equation}\label{eq:angular_distance}
    \cos \Delta \sigma
    = \cos \theta_{\rm f} \cos \theta_{\rm obs}+ \sin \theta_{\rm f} \sin \theta_{\rm obs} \cos\Delta \phi_{\rm f},
\end{equation}
where $\theta_{\rm f}$ and $\phi_{\rm f}$ are the polar and azimuthal angles of the trajectory at $r_{\rm obs}$, and $\Delta \phi_{\rm f} = \phi_{\rm f} - \phi_{\rm obs}$. However, we note that in the numerical implementation we use the more accurate haversine formula for small $\Delta \sigma$~\cite{Sinnott_1984}. A trajectory is considered to intersect with the observer if $\Delta \sigma \rightarrow 0$ and concretely we enforce that $\Delta\sigma \leq 10^{-12}$. Given the nature of the \ac{GSHE}, the initial directions of the \ac{GSHE} trajectories at neighboring $\epsilon$ should lie sufficiently close to each other (or to the initial geodesic direction). Therefore, we typically begin by solving for the initial direction at the highest value of $\epsilon$ that connects the source and observer, then we solve for the $2$\textsuperscript{nd} highest value of $\epsilon$ in a restricted region of the former initial direction and repeat this process down to the smallest $\epsilon$ and the geodesic initial direction.

We first evaluate the \acp{ODE} symbolically in \texttt{Mathematica}~\cite{Mathematica}, expressing them explicitly in the Boyer-Lindquist coordinates. We then export the symbolic expressions to \texttt{Julia}~\citep{Julia} and use the \texttt{DifferentialEquations.jl}~\citep{DifferentialEquations} along with \texttt{Optim.jl} package~\cite{Optim} to integrate the ODEs and optimize the initial conditions, respectively. The Jacobian in~\cref{eq:trajectoryjacobian} is calculated using automatic differentiation implemented in~\texttt{ForwardDiff.jl}~\citep{ForwardDiff}.

%%%%%%%%%%%%%%%%%%%%%%%%%%%%%%%%%%%%%%%%%%%%%%%%%%%%%%%%%%%%%%%%%%%%%%%%

\subsection{Quantifying the time delay}\label{sec:quant_delay}

We write the observer proper time of arrival of a \ac{GSHE} trajectory emitted at coordinate time $t=0$ belonging to the \textit{n}\textsuperscript{th} bundle as $\tau^{\left(n\right)}_{\rm GSHE}(\epsilon, s)$. We specifically denote the proper time of arrival of the geodesic belonging to the \textit{n}\textsuperscript{th} bundle as $\tau_{\rm GO}^{\left(n\right)}$, as it corresponds to the \ac{GO} limit of infinite frequency. We note that
\begin{equation}
    \lim_{\epsilon \rightarrow 0}\tau^{\left(n\right)}_{\rm GSHE}(\epsilon, s) = \tau_{\rm GO}^{\left(n\right)}.
\end{equation}
We will calculate the dispersive \ac{GSHE}-to-geodesic time delay as
\begin{equation}\label{eq:gshe_to_geodesic_delay}
    \Delta \tau^{\left(n\right)} (\epsilon, s) = \tau^{\left(n\right)}_{\rm GSHE}(\epsilon, s) - \tau^{\left(n\right)}_{\rm GO}.
\end{equation}
Additionally, we will also explicitly investigate the birefringent delay between the right and left polarization states
\begin{equation}\label{eq:gshe_to_gshe_delay}
    \Delta \tau^{\left(n\right)}_{\rm R-L}(\epsilon)
    =
    \tau^{\left(n\right)}_{\rm GSHE}(\epsilon, s=+2) - \tau^{\left(n\right)}_{\rm GSHE}(\epsilon, s=-2).
\end{equation}

Having fixed the background Kerr metric mass $M$, or equivalently its Schwarzschild radius $R_{\rm s}$, $\epsilon$ is inversely proportional to the wave packet's frequency $f$. Therefore, the aforementioned time delays can be expressed directly as a function of $f$. Dimension-full units of time can be restored by multiplying the resulting expression by $R_{\rm s} / 2 c$.

%%%%%%%%%%%%%%%%%%%%%%%%%%%%%%%%%%%%%%%%%%%%%%%%%%%%%%%%%%%%%%%%%%%%%%%%

\subsection{Waveform modelling}\label{sec:waveform_modelling}

Due to the frequency- and polarization-dependent observer proper time of arrival delay with respect to the \ac{GO} propagation, $\Delta \tau$, the \ac{GSHE} ``delays'' the circular basis frequency components of the original waveform. We write the circular basis frequency-domain unlensed waveform as $\tilde{h}_{0}(f, s)$. With the notation of~\cref{eq:gshe_to_geodesic_delay}, the \ac{GSHE} produces a frequency-domain waveform
\begin{equation}\label{eq:GSHEwaveform_frequency}
\begin{split}
    \tilde{h}_{\rm GSHE}(f, s) =
    \sum_n e^{-2\pi i f \tau^{\left(n\right)}_{\rm GSHE}(f, s) } \sqrt{\left|\mu^{(n)}(f, s)\right|} \tilde{h}_{0} (f, s).
\end{split}
\end{equation}
The sum runs over the different images, i.e. bundles connecting the source and observer. The exponential encodes the frequency- and polarization-dependent time delay, and the square root encodes the magnification-induced amplitude scaling.

We generate the unlensed linear basis waveform in \texttt{PyCBC}~\cite{PyCBC}, which can equivalently be described in the circular basis. The right and left circularly polarized basis vectors, $e_{\rm R}$ and $e_{\rm L}$, can be related to the plus and cross linearly polarized basis vectors, $e_{+}$ and $e_{\times}$, as
\begin{subequations}\label{eq:linear_to_circular}
\begin{align}
    e_{\rm R} &= \frac{1}{\sqrt{2}}\left(e_{+} + i e_{\times}\right),\\
    e_{\rm L} &= \frac{1}{\sqrt{2}}\left(e_{+} - i e_{\times}\right),
\end{align}
\end{subequations}
discussed, e.g. in Ref.~\cite{MTW}.

As usual, a waveform can be inverse Fourier transformed into the time domain,
\begin{equation}
    h(\tau) = \int\dd f ~ \tilde{h}(f) e^{-2\pi i f \tau},
\end{equation}
where we use $\tau$ to denote the observer proper time. The waveform and detector sensitivity are typically described in the linearly polarized basis. In it, the detector strain is described as
\begin{equation}\label{eq:detector_strain}
    h(\tau) = F_{\rm +} h_{\rm +}(\tau) + F_{\rm x} h_{\rm x}(\tau),
\end{equation}
where $F_{\rm +}$ and $F_{\rm x}$ is the antenna response function to the plus and cross polarization~\cite{Maggiore_GWbook}. Equivalently, the detector strain can be expressed as a function of the circularly polarized waveforms upon a suitable redefinition of the antenna response function.

Beyond visually comparing the \ac{GSHE}-corrected waveforms with their geodesic counterparts, we also quantify their mismatch for a single bundle connecting the source and observer. The mismatch between two waveforms is minimized over the merger time and phase. We denote the mismatch between $h_{\rm GO}$, the \ac{GO} waveform related to the unlensed waveform by including the \ac{GO} magnification $\mu_{\rm GO}$, and $h_{\rm GSHE}$ as
\begin{equation}\label{eq:mismatch}
    \mathcal{M}
    =
    1 - \argmax_{t_c, \phi_c} \frac{\langle h_{\rm GO}, h_{\rm GSHE}\rangle}{\sqrt{\langle h_{\rm GO}, h_{\rm GO}\rangle \langle h_{\rm GSHE}, h_{\rm GSHE}\rangle}},
\end{equation}
where $t_c,\,\phi_c$ are the coalescence time and phase, respectively. The mismatch depends on the noise-weighted inner product between two waveforms
\begin{equation}\label{eq:inner_product}
    \langle a, b\rangle = \Re \int \frac{\tilde{a}^*(f)\tilde{b}(f)}{S(f)}\dd f,
\end{equation}
where $S$ is the noise spectral density amplitude that is set by choosing a \ac{GW} detector. We assume the noise to be flat across all frequencies, $S(f) = 1$, as was done, e.g., in Ref.~\cite{Williams_2020}.

\begin{figure*}
    \centering
    \begin{minipage}[t]{0.99\columnwidth}
    \includegraphics[width=\columnwidth]{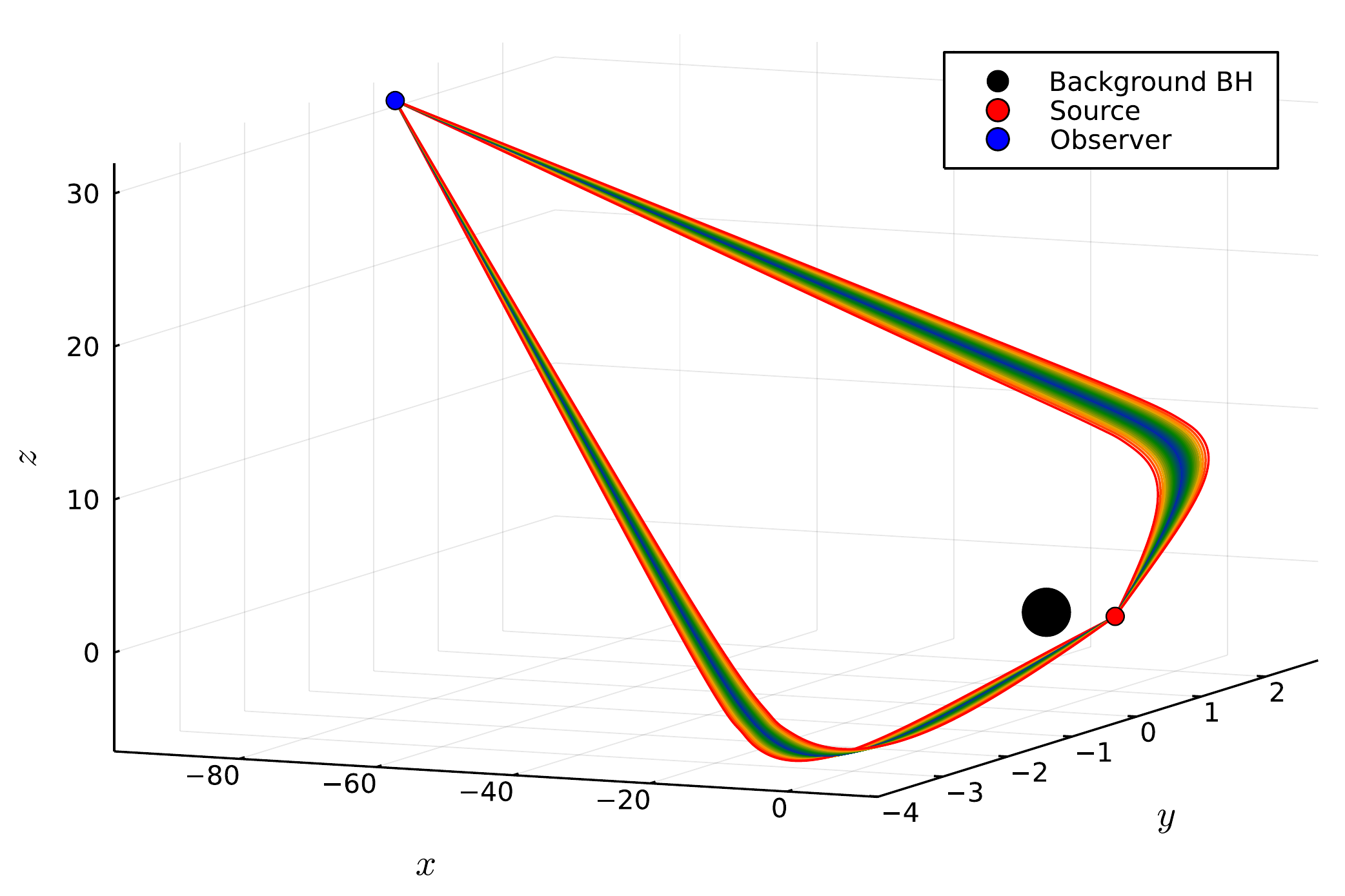}
    \caption{Two bundles of direct trajectories connecting a source at $(5\,R_{\rm s}, \pi/2, 0)$ and an observer at $(50\,R_{\rm s}, 0.4\pi, \pi)$, on the Kerr background metric with $a=0.99$. The \ac{GSHE} trajectories appear as perturbations along their respective geodesic solutions. We plot trajectories with $s = \pm2$ and $10^{-3} \leq \epsilon \leq 10^{-0.3}$, and the units along the space axes are chosen such that $R_{\rm s} = 2$.} \label{fig:example_trajectory}
    \end{minipage}\qquad
    \begin{minipage}[t]{0.99\columnwidth}
    \includegraphics[width=\columnwidth]{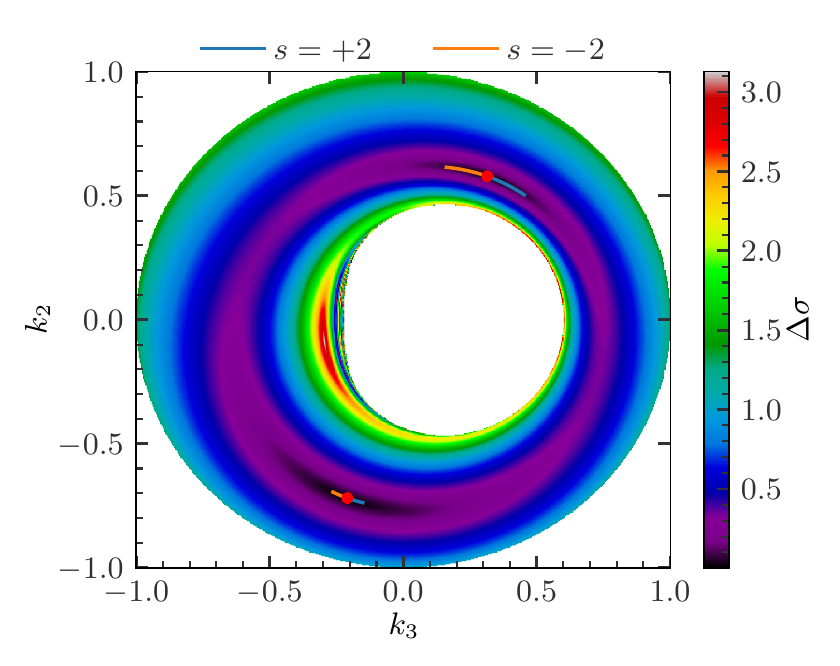}
    \caption{Dependence of the far-sphere angular distance $\Delta\sigma$ on the geodesic initial momentum ($\epsilon = 0$) for a source at $(5\,R_{\rm s}, \pi/2, 0)$, observer at $(50\,R_{\rm s}, 0.4\pi, \pi)$ and $a=0.99$. We minimize $\Delta \sigma$ to find initial momenta that result in connecting trajectories. The highlighted points are the connecting initial geodesic directions, with the neighboring lines showing the $s=\pm 2$ \ac{GSHE} initial directions over $10^{-3} \leq \epsilon \leq 10^{-0.3}$.}
    \label{fig:shadow}
    \end{minipage}
\end{figure*}

For illustration, we now ignore the minimization of the mismatch assuming that the \ac{GSHE} leaves the high-frequency part of the waveform -- the merger -- unchanged
and express the mismatch of a single circular polarization component of a waveform. Furthermore, we assume that $\mu(f,s) = \mu_{\rm GO}$, i.e., that the magnification of the \ac{GSHE} trajectories is equal to the \ac{GO} magnification, which will prove to be a sufficiently good assumption. Because the \ac{GSHE} correction is a phase shift in the frequency domain, we have $\langle h_{\rm GSHE}, h_{\rm GSHE}\rangle = \langle h_{\rm GO}, h_{\rm GO}\rangle$ and
\begin{equation}\label{eq:mismatch_expr}
    \mathcal{M}(h_{\rm GO}, h_{\rm GSHE}, s)
    =
    1 - \frac{\int \dd f |\tilde{h}_{0}(f, s)|^2 \cos \gamma }{\int \dd f  |\tilde{h}_{0}(f, s)|^2},
\end{equation}
where we explicitly wrote the dependence on the circular polarization state, and we define the ``mixing'' angle
\begin{equation}\label{eq:mixingangle}
    \gamma(f, s) = 2\pi f \Delta \tau(f, s).
\end{equation}
Therefore, we have that
\begin{equation}\label{eq:mismatch_apprx}
    \mathcal{M}(h_{\rm GO}, h_{\rm GSHE}, s)
    =
    \frac{1}{2} \frac{\int \dd f ~ \gamma^2 |\tilde{h}_{0}(f, s)|^2}{\int \dd f  |\tilde{h}_{0}(f, s)|^2} + \mathcal{O}(\gamma^4).
\end{equation}
We will demonstrate in~\Cref{sec:numerical_time_delay} that the frequency dependence of $\gamma$ can be isolated from the relevant scaling set by the mutual position of the source and observer, thus further simplifying this expression.

%%%%%%%%%%%%%%%%%%%%%%%%%%%%%%%%%%%%%%%%%%%%%%%%%%%%%%%%%%%%%%%%%%%%%%%%%%%%%%%%

\section{Results}\label{sec:results}

Following the prescription of~\Cref{sec:methodology}, we search for bundles of connecting \ac{GSHE} trajectories between a fixed source and an observer on the Kerr background metric. We investigate how the \ac{GSHE}-induced time delay depends on the mutual position of the source and observer. We discover that in all cases the time delay can be well approximated as a frequency-dependent power law and that the signature of the \ac{GSHE} is a frequency-dependent phase shift in the inspiral part of the observed waveform.

For each configuration, we find the initial directions of a bundle of trajectories by minimizing the angular distance $\Delta \sigma$ of~\cref{eq:angular_distance}. Typically, we search the range $10^{-3} \leq \epsilon \leq 10^{-1}$, with $30$ logarithmically spaced $\epsilon$ values. Everywhere but in~\Cref{sec:directiondependence} we resort to studying only the directly connecting bundles of trajectories to simplify the interpretation. As an example, in~\cref{fig:example_trajectory} we show two directly connecting bundles. The \ac{GSHE} trajectories appear as small deviations from the geodesic trajectories with fixed boundary conditions.

In~\cref{fig:shadow}, we plot an example dependence of $\Delta \sigma$ on the initial ingoing geodesic direction. We minimize $\Delta\sigma$ to find the initial directions that result in a connecting trajectory between a source and an observer. The empty central region indicates the initial directions that penetrate the \ac{BH} horizon, delineating the \ac{BH} shadow. We also overplot in~\cref{fig:shadow} the \ac{GSHE} initial directions upon increasing $\epsilon$ for $s = \pm 2$. If $\epsilon \rightarrow 0$ the \ac{GSHE} initial direction coincides with the initial geodesic direction, otherwise it is twisted by an angle proportional to $\epsilon$.

Now we first characterize the frequency and polarization dependence of the time delay on the system configuration in~\Cref{sec:numerical_time_delay} and then address its impact on the observed waveform in~\Cref{sec:waveform_comparison}.

\begin{figure*}
    \centering
    \includegraphics[width=\textwidth]{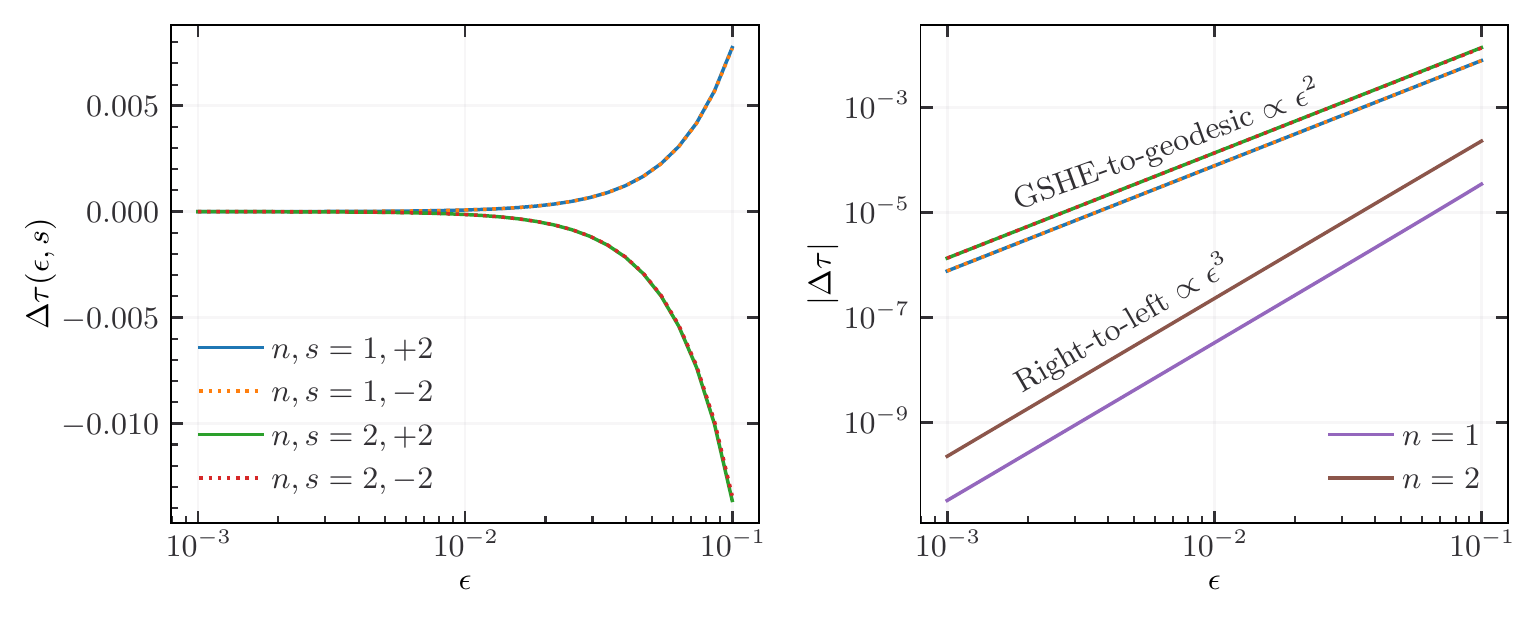}
    \caption{The dispersive \ac{GSHE}-to-geodesic delay with trajectory bundles indexed by $n$ (\emph{left} panel) and the logarithm of the absolute value of the \ac{GSHE}-to-geodesic delay along with the right-to-left delay for each bundle (\emph{right} panel) displaying the power law dependence of the delay. The source is at $(2\,R_{\rm s}, \pi/2, 0)$, observer at $(50\,R_{\rm s}, 0.4\pi, \pi)$ and $a=0.99$.}
    \label{fig:configuration_time_delay}
\end{figure*}

%%%%%%%%%%%%%%%%%%%%%%%%%%%%%%%%%%%%%%%%%%%%%%%%%%%%%%%%%%%%%%%%%%%%%%%%%%%%%%%%

\subsection{Time delay}\label{sec:numerical_time_delay}

In~\cref{fig:configuration_time_delay} we plot the \ac{GSHE}-to-geodesic, $\Delta \tau(\epsilon, s)$, and the right-to-left, $\Delta \tau_{\rm R-L} (\epsilon)$, time of arrival delays for a particular source-observer configuration. We find that, independent of the mutual positions of the source and observer, both $\Delta \tau(\epsilon, s)$ and $\Delta \tau_{\rm R-L}(\epsilon)$ are well described by a power law. Therefore, we introduce
\begin{subequations}\label{eq:coordtime_delay}
\begin{align}
    \Delta \tau(\epsilon, s)
    &\approx
    \beta \epsilon^\alpha,\\
    \Delta \tau_{\rm R-L}(\epsilon)
    &\approx
    \beta_{\rm R-L} \epsilon^{\alpha_{\rm R-L}},
\end{align}
\end{subequations}
for the dispersive \ac{GSHE}-to-geodesic and birefringent right-to-left delay, respectively. In all cases, we find $\alpha \approx 2$ and $\alpha_{\rm R-L} \approx 3$. We note that in the former case both $\alpha$ and $\beta$ have what will turn out to be only a weak dependence on the circular polarization state. The difference between the right and left polarization results in the subdominant, but nonzero, $\Delta \tau_{\rm R-L}(\epsilon)$ delay.

The $\epsilon^2$ dependence of the \ac{GSHE}-to-geodesic delay can be understood as follows. First, the \ac{GSHE} correction to the equations of motion is proportional to $\epsilon$ and, second, to reach the same observer, the \ac{GSHE} initial direction must be rotated with respect to the geodesic initial direction (see the small lines in~\cref{fig:shadow}). The magnitude of this rotation is proportional to $\epsilon$, therefore, altogether these two effects yield an approximate $\epsilon^2$ dependence. The right-to-left delay is a comparison of two perturbed solutions, which produces an $\epsilon^3$ dependence.

On the other hand, the proportionality factors, $\beta$ or $\beta_{\rm R-L}$, are set by the relative position of the source and observer and the \ac{BH} spin. $\beta$ also contains information on the polarization state of the \ac{GW}. As shown in the left panel of~\cref{fig:configuration_time_delay}, in the case of two directly connecting bundles, one of the bundles' \ac{GSHE} trajectories (regardless of the polarization state) arrive with a positive time delay with respect to its geodesic time of arrival, while the other bundles' \ac{GSHE} trajectories arrive with a negative time delay. We verify that this holds in all configurations that we tested.

We may express $\Delta \tau$ explicitly as a function of frequency in dimension-full units of as
\begin{equation}\label{eq:time-delay-dependence}
    \Delta \tau
    \approx
    \beta \left(\frac{2c}{R_{\rm s}}\frac{1}{f}\right)^{\alpha - 1} \frac{1}{f},
\end{equation}
with a similar expression for the right-to-left delay $\Delta \tau_{\rm R-L}$. Thus, the right-to-left delay is suppressed relative to the \ac{GSHE}-to-geodesic delay by an additional power of $2 c / (R_{\rm s} f)$ and generally we have $|\beta| \gg |\beta_{\rm R-L}|$ (exemplified in~\cref{fig:configuration_time_delay}).

Numerically, we find that the \ac{GSHE} trajectories have a ``blind spot'' approximately on the opposite side of the \ac{BH} that cannot be reached, regardless of the initial emission direction. In other words, given a source close to the \ac{BH}, there are spacetime points on a sphere of large $r$ that cannot be reached by \ac{GSHE} trajectories, while these points can be reached by geodesics. The location and size of the blind spot depend on the position of the source, $\epsilon$ (wavelength), polarization, and the \ac{BH} spin. In the Schwarzschild metric, the blind spot is a cone whose size is $\sim 0.5~\mathrm{degrees}$ for $r_{\rm src} = 5\,R_{\rm s}$ and $\epsilon = 0.1$ (upper limit considered in this work). The size decreases with higher $r_{\rm src}$ and lower $\epsilon$, approaching zero when $\epsilon \rightarrow 0$ as there is no blind spot in the geodesic case. The blind spot is exactly centered on the opposite side of the \ac{BH} in the Schwarzschild metric. For a source in the equatorial plane, increasing the \ac{BH} spin slightly tilts the blind spot away from the equatorial plane, and its size remains approximately unchanged. We note that the presence of the blind spot is not a numerical defect and is instead a consequence of the \ac{GSHE} equations. We verify this by inspecting where the \ac{GSHE} trajectories intersect the far-observer sphere upon emission in all possible directions from the source and increasing the numerical accuracy. We leave a further investigation and discussion of the blind spot for future work.

We note that each of the main \ac{GSHE} trajectory bundles has opposite signs of the time delay, cf. Fig. \ref{fig:configuration_time_delay}. The first image to be received has $\beta>0$ (i.e. low frequencies delayed w.r.t. geodesic), while the second image has $\beta<0$ (low frequencies advanced w.r.t. geodesic). As geodesics correspond to extrema of the time delay, we interpret this property as the first bundle being deformed by the \ac{GSHE} into longer time delays, while the second bundle gets distorted in a way that decreases the travel time. This is analogous to standard lensing theory, where images form at extrema of the time-delay function. For a point lens, the first image corresponds to the absolute minimum and the second to a saddle point of the time delay. Angular deformations around the saddle point (as found in Fig. \ref{fig:shadow}) drive the time delay closer to the global minimum, explaining the lower time delay associated with $\beta<0$. The second \ac{GSHE} bundle has negative parity ($\mu<0$), which is consistent with a saddle-point image in the point-lens analogy.

\begin{figure*}
    \centering
    \includegraphics[width=\textwidth]{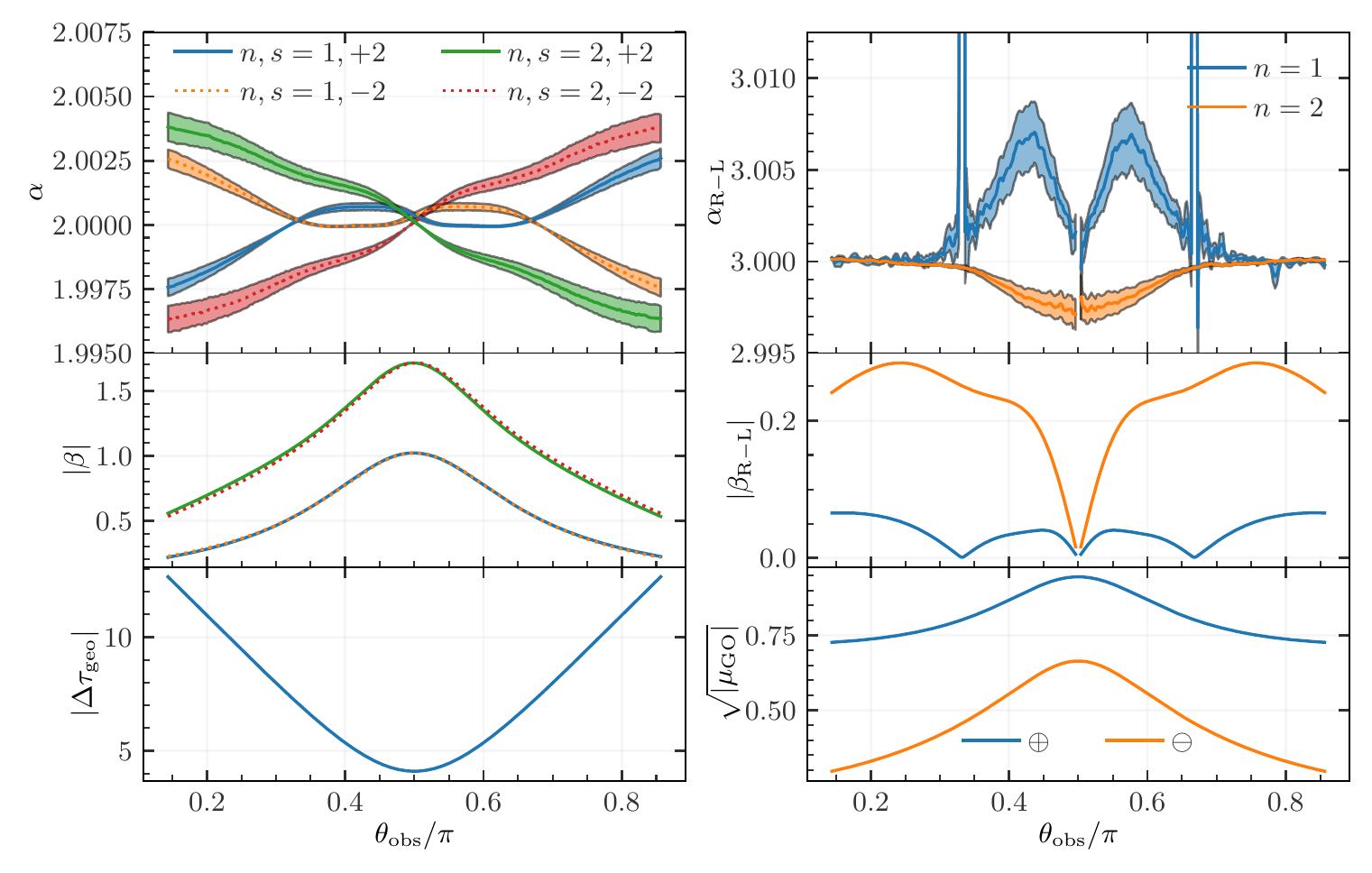}
    \caption{Time delay parametrization upon varying the polar angle of the observer $\theta_{\rm obs}$. The \emph{top} row shows the power law exponent of the dispersive \ac{GSHE}-to-geodesic delay $\alpha$ and of the birefringent right-to-left delay $\alpha_{\rm R-L}$. The \emph{middle} row shows the corresponding power law proportionality factors $\beta$ and $\beta_{\rm R-L}$. The \emph{bottom} row shows the temporal spacing of the two bundles' geodesics $\Delta \tau_{\rm geo}$ and the geodesic magnification $\mu_{\rm GO}$ ($\oplus$ and $\ominus$ indicate positive and negative parity, respectively). The source is otherwise at $(2\,R_{\rm s}, \pi/2, 0) $, observer at $(50\,R_{\rm s}, \theta_{\rm obs}, \pi) $ and $a = 0.99$. When both the source and observer are in the equatorial plane the right-to-left delay vanishes due to reflection symmetry. $\Delta \tau_{\rm geo}$ is nonzero and $\mu_{\rm GO}$ remains finite when $\theta_{\rm obs} = \pi/2$ because of the \ac{BH} spin.
    }
    \label{fig:thetaobspowerlaw_fit}
\end{figure*}

\begin{figure*}
    \centering
    \includegraphics[width=\textwidth]{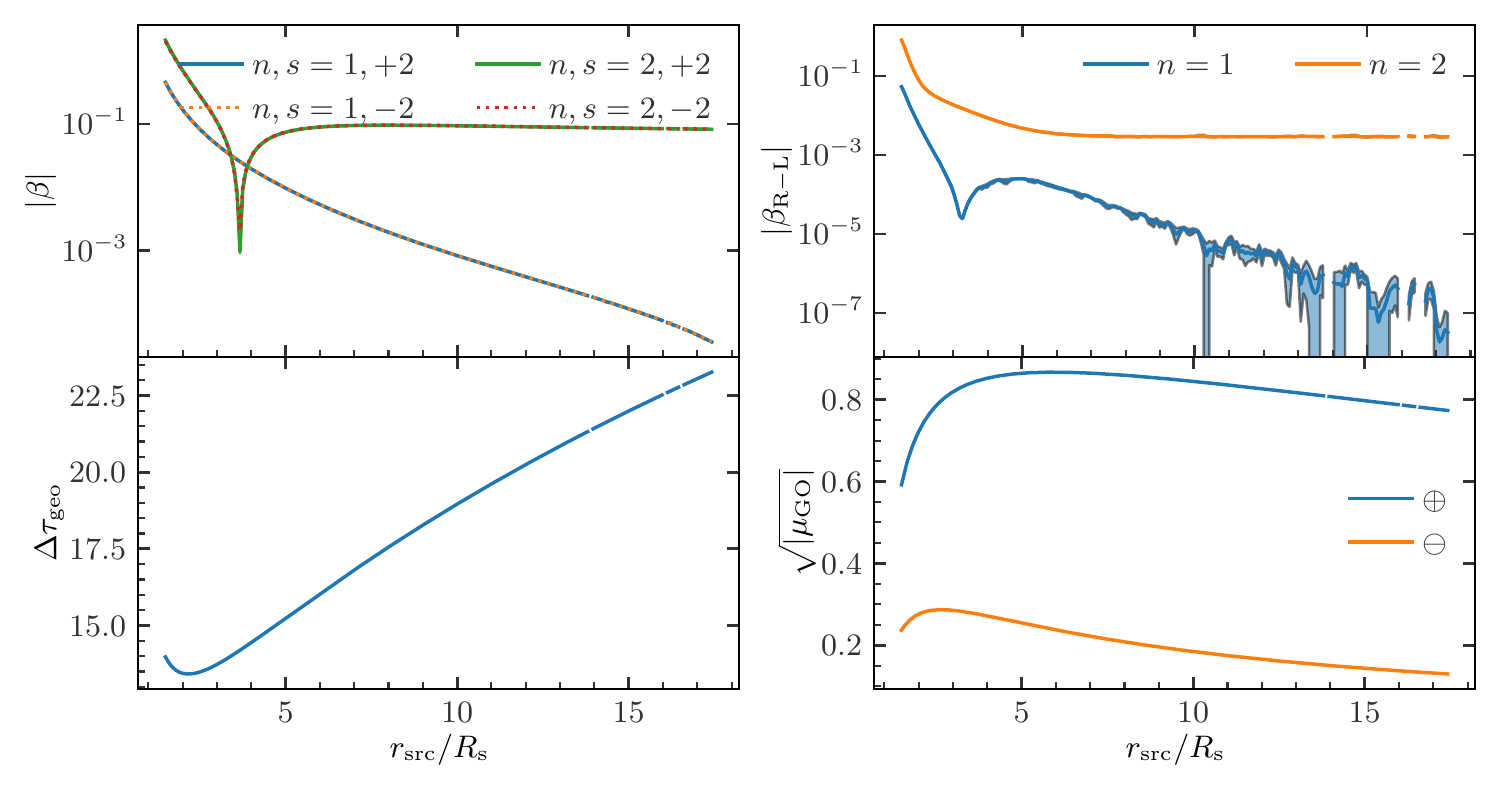}
    \caption{Time delay parametrization upon varying the source radial distance $r_{\rm src}$. Similarly to~\cref{fig:thetaobspowerlaw_fit}, the \emph{top} row shows $\beta$ and $\beta_{\rm R-L}$. The \emph{bottom} row shows $\Delta \tau_{\rm geo}$ and $\mu_{\rm GO}$. The source is otherwise at $(r_{\rm src}, \pi/2, 0) $, observer at $(50\,R_{\rm s}, 0.4\pi, 0.75\pi) $ and $a=0.99$. We have that $\alpha \approx 2$ and $\alpha_{\rm R-L} \approx 3$. The $n = 1$ bundle completes an azimuthal angle of $5\pi/4$ and is deflected in the strong-field regime of the \ac{BH}. Consequently, $\beta$ approaches a constant value, however this bundle is exponentially demagnified.}
    \label{fig:radial_powerlaw_fit}
\end{figure*}

We now describe the dependence of the time delay on the mutual position of the source and observer and on the spin of the \ac{BH}. The \ac{BH} mass enters only when we relate $\epsilon$ to frequency and restore dimension-full units of time. To demonstrate the dependence, we vary the observer's polar angle $\theta_{\rm obs}$ and the radial distance $r_{\rm src}$ of the source from the \ac{BH}. We also study the directional dependence of the \ac{GSHE}, wherein we keep the source fixed but calculate the delay as a function of the emission direction. Additionally, the variation of the \ac{BH} spin and observer polar angle is discussed in~\cref{sec:more_timedelay}. In all cases, we place the observer at $r_{\rm obs} = 50\,R_{\rm s}$ after verifying that the time delay becomes approximately independent of $r_{\rm obs}$ once the observer is sufficiently far away. When we plot the power law parameters describing the time delay, we include the $1\sigma$ error bars estimated by bootstrapping. Upon varying the location of the source or observer, we associate bundles by similarity in time of arrival and initial direction.

%%%%%%%%%%%%%%%%%%%%%%%%%%%%%%%%%%%%%%%%%%%%%%%%%%%%%%%%%%%%%%%%%%%%%%%%%%%%%%%%

\subsubsection{Dependence on the observer polar angle}

We begin by showing the dependence of the power law parameters, describing the time delay, on $\theta_{\rm obs}$ in~\cref{fig:thetaobspowerlaw_fit}. We only consider direct bundles (i.e., no complete loops around the \ac{BH}) indexed by $n$. The source is kept at $(2\,R_{\rm s}, \pi/2, 0)$, observer at $(50\,R_{\rm s}, \theta_{\rm obs}, \pi)$ and $a=0.99$. In all cases, we find near perfect agreement with the power law parameterized as in~\cref{eq:coordtime_delay}, according to $\alpha \approx 2$ and $\alpha_{\rm R-L}\approx 3$. The power law proportionality of the \ac{GSHE}-to-geodesic delay is typically close to an order of magnitude larger than that of the right-to-left delay, in agreement with the example configuration shown in~\cref{fig:configuration_time_delay}. While the \ac{GSHE}-to-geodesic delay is maximized when both source and observer are located in the equatorial plane, the right-to-left delay is zero in such a case, because of the reflection symmetry about the equatorial plane. We numerically verify that this condition applies more generally whenever $\theta_{\rm obs} + \theta_{\rm src} = \pi$.

Furthermore, in the bottom panels of~\cref{fig:thetaobspowerlaw_fit} we plot $\Delta \tau_{\rm geo}$ defined as
\begin{equation}
    \Delta \tau_{\rm geo} = \tau_{\rm GO}^{(n=1)} - \tau_{\rm GO}^{(n=2)}.
\end{equation}
This is the \ac{GO} time of arrival difference between the geodesics of the two direct bundles indexed by $n=1,2$. As expected, $\Delta \tau_{\rm geo}$ is symmetric about $\theta_{\rm obs} = \pi/2$ as the source is in the equatorial plane. In all cases, the temporal spacing of the directly connecting bundles is several orders of magnitude larger than the \ac{GSHE}-induced delay within a single bundle. In the second bottom panel we show $\mu_{\rm GO}$, the magnification factor of the geodesic trajectory of each of the two bundles, which shows a weak dependence on $\theta_{\rm obs}$. The magnification factor is unique for each trajectory in the bundle and therefore is also a function of $\epsilon s$. However, we find that its dependence on $\epsilon s$ is negligible, and therefore we only plot the geodesic magnification factor. In fact, it will turn out that in all cases considered in this work the $\epsilon s$ dependence of the magnification is negligible and we may write that
\begin{equation}
    \mu(f, s) \approx \mu_{\rm GO}.
\end{equation}
Similarly, we find that in all cases the $\epsilon s$ dependence of the gravitational redshift, discussed in~\cref{eq:gravredshift}, is negligible and well described by the gravitational redshift of the geodesic trajectory. In all cases, the image from one bundle has positive parity and negative parity for the other bundle, which also consistently holds when varying $\theta_{\rm obs}$.

\begin{figure*}
    \centering
    \begin{minipage}[t]{0.99\columnwidth}
    \includegraphics[width=\columnwidth]{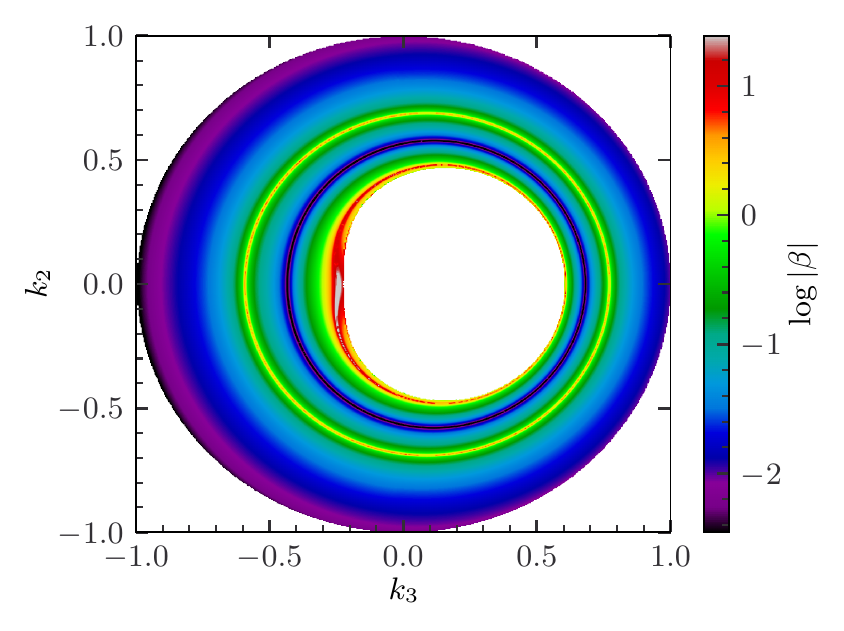}
        \caption{The dispersive \ac{GSHE}-to-geodesic delay parameter $\beta$ as a function of the maximum $\epsilon = 0.01$ initial momentum parameterized by $k_2$ and $k_3$ (\cref{eq:shadow_initial_momentum}). The source is placed at $(5\,R_{\rm s}, \pi/2, 0) $ and the ``observer'' is defined as the point where the $\epsilon_{max}$ trajectory intersects a sphere of radius $50\,R_{\rm s}$. Each pixel represents an $\epsilon$ bundle of trajectories.}
        \label{fig:shadow_beta}
    \end{minipage}\qquad
    \begin{minipage}[t]{0.99\columnwidth}
    \includegraphics[width=\columnwidth]{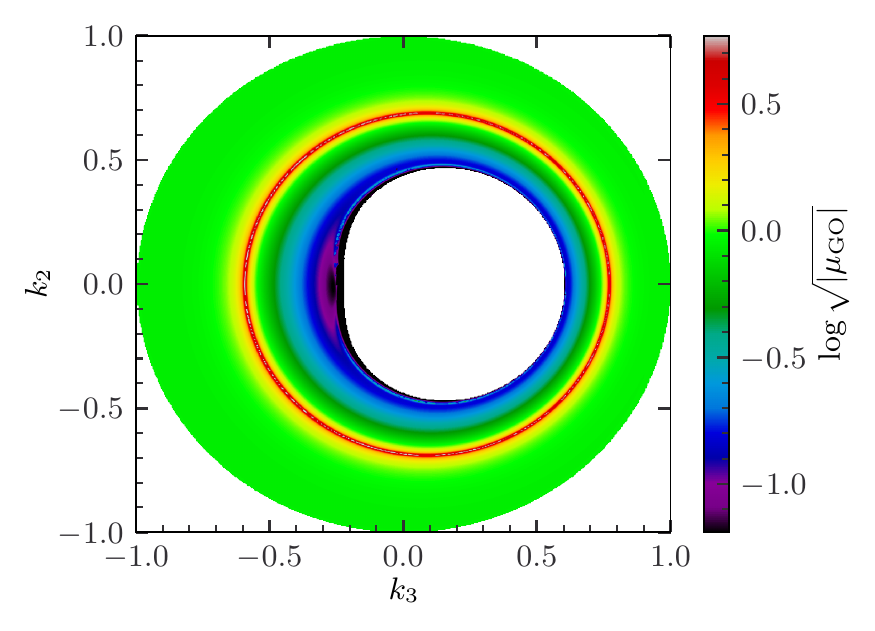}
        \caption{The geodesic magnification $\mu_{\rm GO}$ as a function of the initial emission direction $(k_2, k_3)$, corresponding to the $\beta$ calculation of~\cref{fig:shadow_beta}. The source is placed at $(5\,R_{\rm s}, \pi/2, 0) $. The outer green ring of~\cref{fig:shadow_beta} is magnified (red ring), while trajectories passing close to the \ac{BH} shadow are demagnified.}
        \label{fig:shadow_mu}
    \end{minipage}
\end{figure*}

%%%%%%%%%%%%%%%%%%%%%%%%%%%%%%%%%%%%%%%%%%%%%%%%%%%%%%%%%%%%%%%%%%%%%%%%%%%%%%%%

\subsubsection{Dependence on the source radial distance}

In~\cref{fig:radial_powerlaw_fit}, we plot $\beta$, $\beta_{\rm R-L}$, $\Delta \tau_{\rm geo}$ and $\mu_{\rm GO}$ when varying $r_{\rm src}$. We do not explicitly show the power law exponent. However, we verify that $\alpha \approx 2$ and $\alpha_{\rm R-L} \approx 3$ remain satisfied. The source is at $(r_{\rm src}, \pi/2, 0) $, the observer is at $(50\,R_{\rm s}, 0.4\pi, 3\pi/4) $ and $a=0.99$. We do not place the observer directly opposite the source, instead choosing $\phi_{\rm obs} = 3\pi / 4$. This ensures that one of the bundles completes an azimuthal angle of $3\pi/4$, while the other $5\pi/4$. When the source is moved further away from the \ac{BH} the former will propagate directly to the observer without entering the strong-field regime of the \ac{BH}, whereas the latter is forced to effectively sling by the \ac{BH}.

Figure \ref{fig:radial_powerlaw_fit} shows that in the case of direct propagation, both $\beta$ and $\beta_{\rm R-L}$ decay exponentially as the trajectories do not experience strong gradients of the gravitational field, for example approximately $\beta \propto 10^{-0.2 r_{\rm src}/ R_{\rm s}}$. On the other hand, when the trajectories are forced to sling around the \ac{BH}, we find that both $\beta$ and $\beta_{\rm R-L}$ tend to a constant, non-negligible value since regardless of how distant the source is, the trajectories pass close to the \ac{BH}. This suggests that it is possible to place the source far away from the \ac{BH} and still obtain strong \ac{GSHE} corrections, provided that the trajectories pass by the \ac{BH} as expected in strong lensing.

In the bottom left panel, we plot the temporal spacing of the two bundles, $\Delta \tau_{\rm geo}$, which is proportional to $r_{\rm src}$. In the bottom right panel, we plot the absolute value of $\mu_{\rm GO}$. Just as before, the dependence of both magnification and gravitational redshift on $\epsilon s$ is negligible. We previously noted that for the bundle that is forced to sling around the \ac{BH} we obtain a \ac{GSHE} correction that is approximately independent of $r_{\rm src}$. However, this bundle is also exponentially demagnified, as shown in~\cref{fig:radial_powerlaw_fit}, with approximately $\mu_{\rm GO} \propto 10^{-0.05 r_{\rm src} / R_{\rm s}}$. Since it is the square root of the magnification that scales the signal, despite the exponential demagnification, this configuration remains an interesting avenue for detecting the \ac{GSHE}, as long as $r_{\rm src}$ is not too large.

%%%%%%%%%%%%%%%%%%%%%%%%%%%%%%%%%%%%%%%%%%%%%%%%%%%%%%%%%%%%%%%%%%%%%%%%%%%%%%%%

\subsubsection{Directional dependence of the \ac{GSHE}}\label{sec:directiondependence}

We now report on the directional dependence of the time delay from the source point of view, considering trajectories that initially point towards the \ac{BH}. We emit a \ac{GSHE} trajectory from the source at the maximum value of $\epsilon$ in the direction parameterized by $(k_2, k_3)$, introduced in~\cref{eq:shadow_initial_momentum}. Then we record the angular coordinates where this trajectory intersects a far origin-centered sphere of radius $50\,R_{\rm s}$, setting that location as the ``observer'' for the above choice of initial direction.  We find the remaining \ac{GSHE} and geodesic trajectories that connect to the same point and form a bundle of trajectories. Starting with the maximum value of $\epsilon$ guarantees that we never fix an observer in the blind spot of any \ac{GSHE} trajectories.

We characterize each bundle belonging to an initial choice of $(k_2, k_3)$ by $\beta$ of the right-polarized rays in the left panel of~\cref{fig:shadow_beta} (note that the directions in this figure correspond to the initial directions of the \ac{GSHE} rays with maximum $\epsilon = 0.01$). Throughout, we keep the source at $(5\,R_{\rm s}, \pi/2, 0)$ and do not calculate the left-polarized rays, as those behave sufficiently similarly. This time, we do not eliminate the initial directions that result in trajectories that completely loop around the \ac{BH}. We still have $\alpha \approx 2$, although a small fraction of the initial directions, particularly close to the \ac{BH} horizon, deviate by $\sim1\%$. The left panel of~\cref{fig:shadow_beta} shows a characteristic ring of initial directions that produce $|\beta| \sim 1$, which approximately corresponds to the trajectories that are mapped to the point opposite side of the \ac{BH} (more precisely, these trajectories are mapped close to the edge of the blind spot for the maximum $\epsilon = 0.01$). The initial directions close to the \ac{BH} horizon produce $|\beta| \sim 10$, although these are extreme configurations that completely loop around the \ac{BH} and are demagnified. The initial directions of the outgoing trajectories (not shown in~\cref{fig:shadow_beta}) result in $|\beta|$ lower than the minimum of the ingoing trajectories and therefore are of little interest for the detection of the \ac{GSHE}.

\begin{figure*}
    \centering
    \begin{minipage}[t]{0.99\columnwidth}
    \includegraphics[width=\columnwidth]{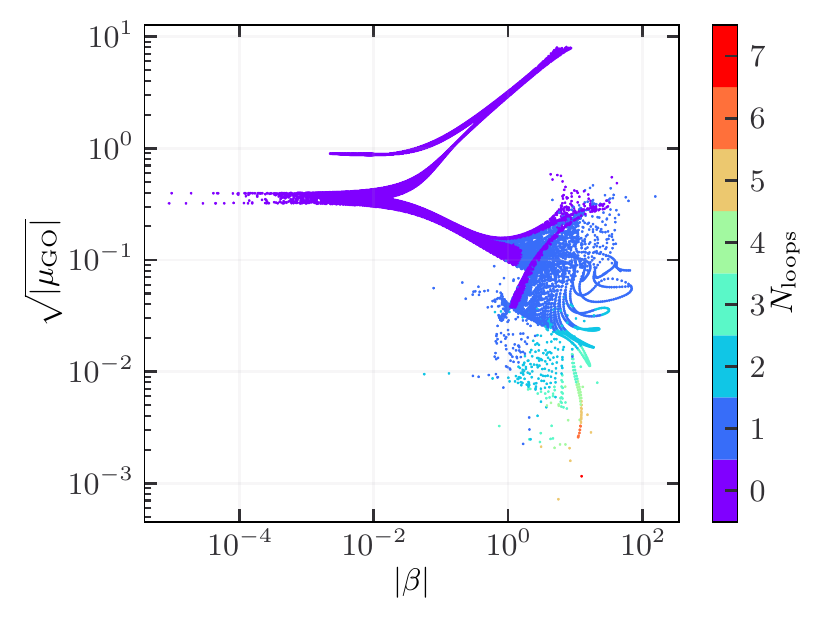}
        \caption{The relation between the geodesic magnification $\mu_{\rm GO}$ and $|\beta|$ in pixels of~\cref{fig:shadow_beta}. The colour represents the number of complete loops $N_{\rm loops}$ around the \ac{BH}. The magnified region in the top right consists of the high $|\beta|$ outer green ring in~\cref{fig:shadow_beta}. The demagnified region in the bottom right consists of bundles that pass very close to the \ac{BH} horizon.}
        \label{fig:beta2mu_scatter}
    \end{minipage}\qquad
    \begin{minipage}[t]{0.99\columnwidth}
    \includegraphics[width=\columnwidth]{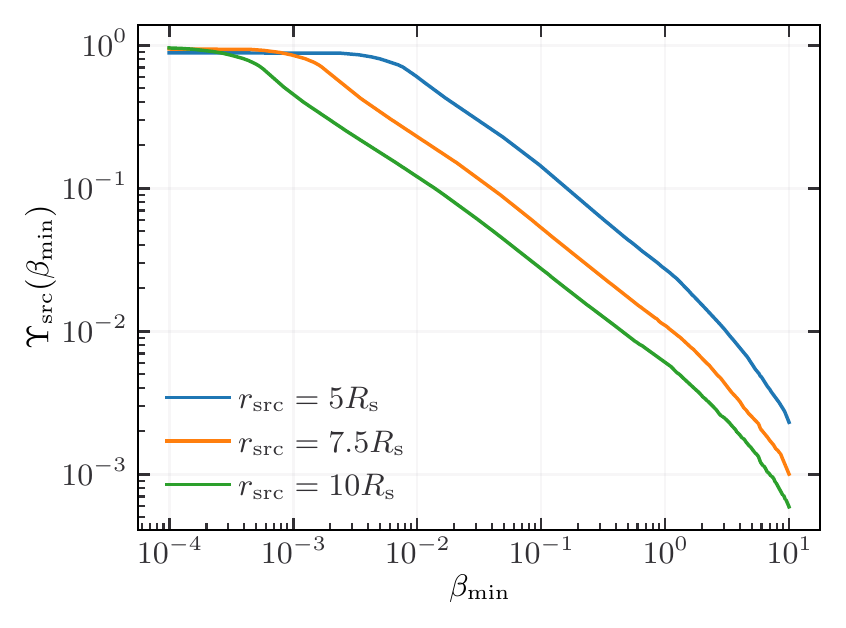}
        \caption{Fraction of the source celestial half-sphere that yields $|\beta| \geq \beta_{\min}$. In the region where $\Upsilon$ is decaying we approximately have $\Upsilon \propto 1/r_{\rm src}^2$. The source is in the equatorial plane and $a=0.99$. The $r_{\rm src} = 5\,R_{\rm s}$ line corresponds to~\cref{fig:shadow_beta}.}
        \label{fig:upsilon}
    \end{minipage}
\end{figure*}

Having demonstrated how $|\beta|$ depends on the direction of the emission, we now study the dependence of the corresponding magnification factor. We again verify that the deviation of the magnification as a function of $\epsilon$ from its geodesic is at most $\sim 1\%$, although typically smaller by up to several orders of magnitude. In~\cref{fig:shadow_mu}, we show $\mu_{\rm GO}$ as a function of the emission direction, matching~\cref{fig:shadow_beta}. Additionally, in~\cref{fig:beta2mu_scatter} we explicitly show a scatter plot of $\mu_{\rm GO}$ and $\beta$ corresponding to the pixels in~\cref{fig:shadow_beta} and~\cref{fig:shadow_mu}. The scatter plot displays two high $|\beta|$ tails -- one where $|\beta|$ is positively correlated with $\mu_{\rm GO}$ and one where the correlation is negative. The former corresponds to the aforementioned outer green ring of~\cref{fig:shadow_beta} of bundles that approximately reach the point on the other side of the \ac{BH} and are magnified as they converge into a smaller region. The latter is demagnified, as it consists of bundles that pass close to the \ac{BH} horizon and are sensitive to the initial direction. Therefore, it is the outer green ring of~\cref{fig:shadow_beta} that comprises a promising landscape for observing the \ac{GSHE} due to its high $|\beta|$ and $|\mu_{\rm GO}| > 1$.

We calculate the fraction of the source celestial half-sphere of~\cref{fig:shadow_beta} that yields $|\beta| > \beta_{\rm min}$ of the \ac{GSHE}-to-geodesic delay for the right-polarized rays as
\begin{equation}\label{eq:upsilon}
    \Upsilon_{\rm src}(\beta_{\min}) = \frac{1}{2\pi} \int H(|\beta| - \beta_{\min}) \sin\psi \dd \psi \dd \rho,
\end{equation}
where the integral runs over the celestial half-sphere of ingoing trajectories, $H(\cdot)$ is the Heaviside step function defined as $H(x) = 1$ if $x > 0$ and $0$ otherwise. We note that a fraction of the half-sphere is covered by the shadow of the \ac{BH} and, therefore, $\Upsilon_{\rm src}(0) < 1$. We plot $\Upsilon_{\rm src}(\beta_{\min})$ in~\cref{fig:upsilon} for sources at $(r_{\rm src}, \pi/2, 0)$, where we choose $r_{\rm src} = 5, 7.5, 10\,R_{\rm s}$ and $a=0.99$. We find that for $r_{\rm src} = 5\,R_{\rm s}$ about $5\%$ of the ingoing half-sphere yield $|\beta| \gtrsim 0.5$, and we verify that $\Upsilon_{\rm src}$ is approximately proportional to $1/r_{\rm src}^2$ in the region where it is decaying.

Similarly, we calculate the fraction of the far sphere of radius $r = r_{\rm obs}$ where an observer would measure $|\beta| > \beta_{\rm min}$ and $\mu > |\mu_{\rm min}|$:
\begin{equation}\label{eq:upsilon_observer}
\begin{split}
    \Upsilon_{\rm obs} (\beta_{\min}, \mu_{\min})
    =&\frac{1}{4\pi}\int \mathcal{S}(\beta,\mu) \sin \phi \dd \phi \dd \theta, \\
    =& \frac{1}{4\pi} \int \mathcal{S}(\beta,\mu) \frac{\sin \psi}{|\mu(\psi,\rho)|} \dd \psi \dd \rho.
\end{split}
\end{equation}
Here, $(\theta, \phi)$ are coordinates on the spacetime sphere $r = r_{\rm obs}$, and $(\rho, \psi)$ are coordinates on the celestial sphere of the source. The Jacobian relating both coordinates is the inverse of the magnification, as has been included in the second line: this can be intuitively understood as magnified/demagnified trajectories being focused/spread out and therefore less/more likely. The integral is weighted by a selection function
\begin{equation}
 \mathcal{S} =   H(|\beta (\phi, \theta)| - \beta_{\min}) H(|\mu (\phi, \theta)| - \mu_{\min}),
\end{equation}
eliminating trajectories that are either too faint to be detected or for which the \ac{GSHE} is undetectable. We are considering trajectories that loop around the \ac{BH}. Therefore, multiple trajectories can reach an observer, so $\Upsilon_{\rm obs} > 1$ in general when computing probabilities (\cref{sec:disc_detectability}).

\begin{figure*}
    \includegraphics[width=\columnwidth]{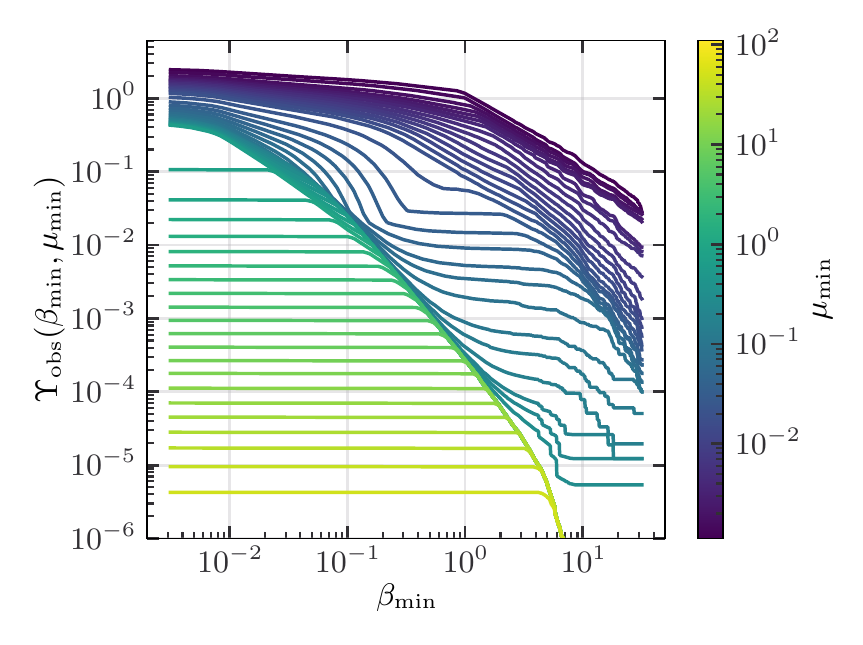}
    \includegraphics[width=\columnwidth]{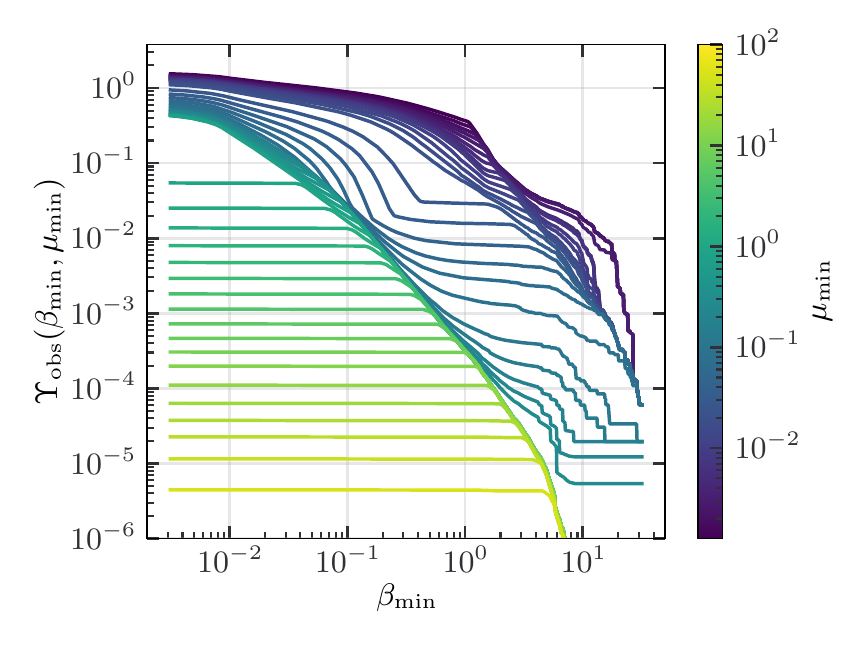}
\caption{Observer's cumulative GSHE probability as a function of the minimum magnification (absolute value), including all trajectories (left) and excluding trajectories that loop around the \ac{BH} (right). Only trajectories with $|\mu|>10^{-3}$ are considered. Differences are appreciable only for $\mu\ll 1$.}\label{fig:Ups_obs}
\end{figure*}

\cref{fig:Ups_obs} shows the observer's cumulative \ac{GSHE} probability for different magnification cuts. Two cases are considered: the left panel allowing for any number of loops around the \ac{BH}, which has a maximum number of $7$ in our numerical exploration. The right panel restrict the results to zero loops, although strongly deflected trajectories with $\alpha < 2\pi$ are still considered (these trajectories could be discriminated by the sign of $\mu$, as they have negative parity). The differences are noticeable only for faint trajectories with $|\mu| \ll 1$: for $\beta_{\min}\lesssim 1$ $\Upsilon_{\rm obs}$ is larger than unity, reflecting the existence of these additional trajectories. For $\beta_{\min}\gtrsim 1$ the additional loops increase the probability considerably. Note that the high $\beta$ end is restricted by the resolution in our numerical exploration.

The results can be adapted to different distances between the source and the \ac{BH} without an additional sampling. The \ac{GSHE} probability for the source scales as $\Upsilon_{\rm src}\propto r_{\rm src}^{-2}$, cf.~\cref{fig:upsilon}, as the regions contributing to the different values of $\beta$ span a smaller portion of the source's sphere. Additionally, the magnification scales by the same factor $\mu \propto r_{\rm src}^{-2}$ \cite{Gondan:2021fpr}, reflecting the divergence of rays before encountering the lens.

Lastly, in~\Cref{sec:parity_sign} we discuss the relation between the image parity of trajectory bundles of~\cref{fig:shadow_beta} and the sign of the \ac{GSHE}-to-geodesic delay.~\Cref{sec:upsilon_details} discusses the effect of multiple loops and sign of $\beta$ on the observer's probability.

%%%%%%%%%%%%%%%%%%%%%%%%%%%%%%%%%%%%%%%%%%%%%%%%%%%%%%%%%%%%%%%%%%%%%%%%%%%%%%%%

\subsubsection{Dependence on the remaining parameters}

We postpone the discussion of varying the \ac{BH} spin $a$ and the azimuthal angle of the observer $\phi_{\rm obs}$ to~\Cref{sec:more_timedelay}. However, we highlight that in the Schwarzschild metric, the right-to-left delay vanishes because of reflection symmetry. On the other hand, the \ac{GSHE}-to-geodesic delay is maximized in Schwarzschild, which we attribute to the fact that lowering the \ac{BH} spin pushes its horizon outwards, and therefore the trajectories pass closer to the \ac{BH} horizon where the gradient of the gravitational field is larger. We verify that this behavior is not a consequence of a particular source-observer configuration and qualitatively holds in general.

%%%%%%%%%%%%%%%%%%%%%%%%%%%%%%%%%%%%%%%%%%%%%%%%%%%%%%%%%%%%%%%%%%%%%%%%%%%%%%%%

\subsection{Waveform comparison}\label{sec:waveform_comparison}

We consider the \texttt{IMRPhenomXP} waveform~\cite{IMRPhenomXP} of a $25$ and $10 M_\odot$ binary \ac{BH} merger observed at an inclination angle of $0.9\pi$ with the spin of the primary along the $z$-axis $a_z=0.7$ and $0$ along the remaining axes and zero spin of the secondary. The frequency-domain waveform is generated from $40~\mathrm{Hz}$ to $1024~\mathrm{Hz}$, though the merger frequency is $\sim 225~\mathrm{Hz}$. Following~\cref{eq:epsilon_SI}, we fix the background mass to achieve some maximum value $\epsilon_{\max}$ at the lower frequency limit, since $\epsilon \propto 1/f$.

As an example, for $\epsilon_{\max} = 0.1$ this amounts to $M \sim 5\times10^4~M_\odot$. Following~\cref{eq:time-delay-dependence}, the \ac{GSHE}-to-geodesic and right-to-left observer time delays are
\begin{subequations}
\begin{align}
    \Delta \tau(f)
    &\approx
    3~\mathrm{ms}\,\beta \left(\frac{5\times 10^4 M_\odot}{M}\right) \left(\frac{40~\mathrm{Hz}}{f}\right)^2,\\
    \Delta \tau_{\rm R-L}(f)
    &\approx
    0.3~\mathrm{ms}\,\beta_{\rm R-L} \left(\frac{5\times 10^4 M_\odot}{M}\right)^2 \left(\frac{40~\mathrm{Hz}}{f}\right)^3.
\end{align}
\end{subequations}
The \ac{GSHE}-to-geodesic delay is the dominant component. Moreover, typically $|\beta| \gg |\beta_{\rm R-L}|$ as demonstrated in~\Cref{sec:numerical_time_delay}. The \ac{GSHE}-to-geodesic delay shifts both polarizations in approximately the same direction with respect to the geodesic, as exemplified in~\Cref{fig:configuration_time_delay}. Their difference is the right-to-left delay, which is negligible in most cases. Therefore, we will focus on the difference between the \ac{GSHE}-corrected and geodesic-only waveforms.

\begin{figure*}
    \centering
    \includegraphics[width=\textwidth]{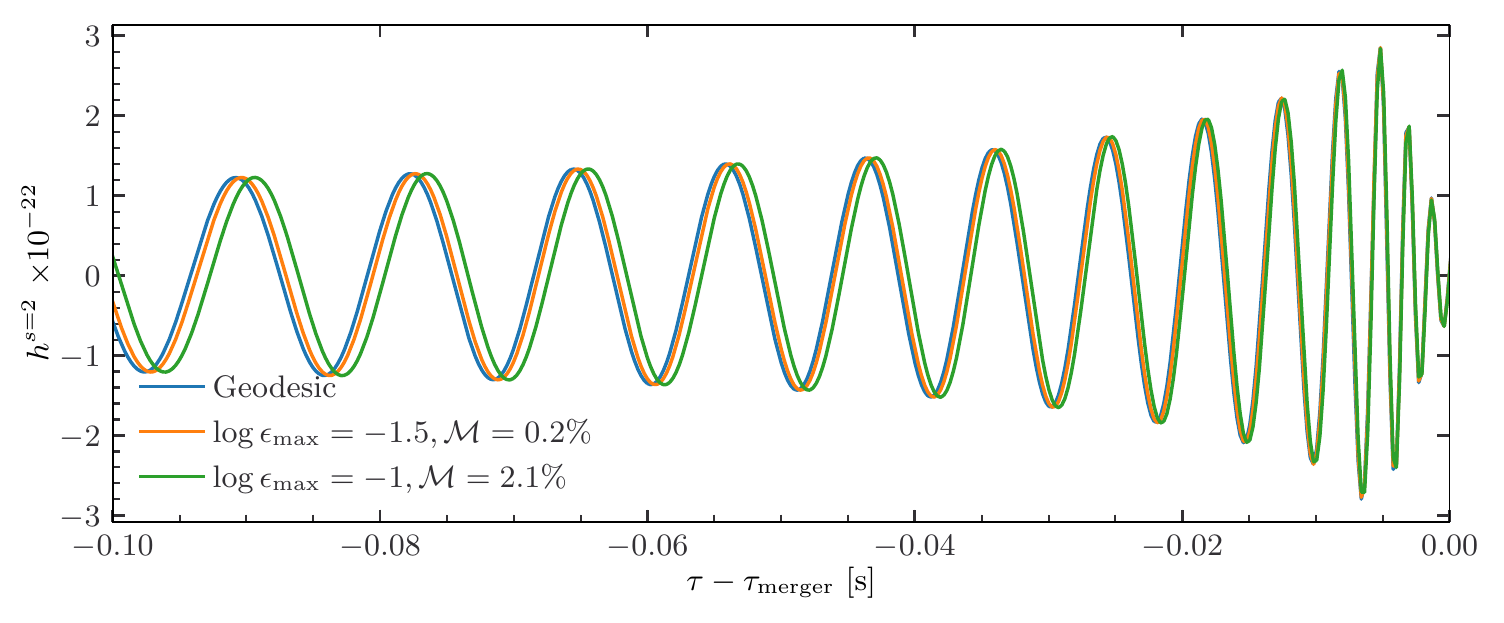}
    \caption{The \ac{GSHE}-corrected and geodesic-only right polarization waveforms of a $25$ and $10M_\odot$ merger if $\beta = 2$. We show two cases of $\epsilon_{\rm max}$, the perturbation strength at the lower frequency range of the waveform, along with the corresponding \ac{GSHE}-induced mismatch. The \ac{GSHE} manifests as a frequency-dependent phase shift in the inspiral part of the signal.}
    \label{fig:circ_waveform}
\end{figure*}

In~\cref{fig:circ_waveform}, we compare the right-polarization geodesic-only and \ac{GSHE}-corrected waveforms for $\beta = 2$ separately if $\log \epsilon_{\max} = -1.5, -1$. This choice of $\beta$ is large enough to demonstrate the \ac{GSHE}, but still reasonably likely, as we showed in~\cref{fig:shadow_beta} and~\cref{fig:upsilon}. We follow the modeling prescription of~\cref{eq:GSHEwaveform_frequency}. Even in the former, more conservative $\epsilon_{\max}$ case, the effect on the waveform is clearly visible and manifested as a frequency-dependent phase shift in the inspiral phase of the merger. This is because the merger and the ringdown are propagated by higher frequency components whose \ac{GSHE} correction is suppressed as $\sim 1/f^2$. Consequently, the intrinsic parameters inferred from the inspiral part of the waveform may appear inconsistent with the merger and ringdown part of the waveform if the \ac{GSHE} is not taken into account. We do not explicitly show the detector strain, which is a linear combination of the right- and left-polarization state waveforms whose phase difference due to the \ac{GSHE} is negligible.

In~\cref{fig:mismatch} we plot the mismatch of the right-polarization waveform calculated following~\cref{eq:mismatch_expr}. We assume that in $\Delta\tau$ the exponent is $\alpha = 2$. We show the mismatch for several choices of $\epsilon_{\max}$, which is equivalent to scaling the background mass $M$ while keeping the waveform fixed. Following~\cref{eq:mismatch_apprx}, this shows that we can approximate the mismatch as $\mathcal{M} \propto \beta^2$ for small mixing angles $\gamma$.

%%%%%%%%%%%%%%%%%%%%%%%%%%%%%%%%%%%%%%%%%%%%%%%%%%%%%%%%%%%%%%%%%%%%%%%%%%%%%%%%

\section{Discussion}\label{sec:discussion}

In the derivation of the \ac{GSHE} and throughout this work, several simplifying assumptions have been made to demonstrate the viability of this effect for future detection. We now first comment on the neglected higher-order contributions to the \ac{GSHE} in~\Cref{sec:disc_higherordercontribs}, the source-observer placement in~\Cref{sec:discussion_source} and the \ac{GW} emission modelling in~\Cref{sec:disc_anisotropy}. Then, in~\Cref{sec:disc_detectability} we discuss the prospects of detecting the \ac{GSHE} and, finally, in~\Cref{sec:disc_tgr} we discuss its relation to tests of \ac{GR} and beyond-\ac{GR} theories.

%%%%%%%%%%%%%%%%%%%%%%%%%%%%%%%%%%%%%%%%%%%%%%%%%%%%%%%%%%%%%%%%%%%%%%%%%%%%%%%%

\subsection{Higher-order GSHE contributions}\label{sec:disc_higherordercontribs}

The \ac{GSHE} equations describe the motion of a wave packet energy centroid and are only valid up to first order in wavelength. The relevant indicator is the \ac{WKB} perturbation parameter $\epsilon$, the ratio between the wave packet wavelength and the \ac{BH} Schwarzschild radius. In the limit of $\epsilon \rightarrow 0$ the geodesic propagation of the wave packet is recovered, while $\epsilon \sim 1$ is the regime of wave-like phenomena, wherein the wavelength is comparable to the characteristic length scale of the system. Going further, if $\epsilon \rightarrow \infty$ we do not expect wave propagation to be significantly affected by the presence of the \ac{BH} as in this limit the presence of the \ac{BH} becomes negligible (see, for example, Ref.~\citep[Fig. 2]{leite2017absorption}).

\begin{figure}
    \centering
    \includegraphics[width=\columnwidth]{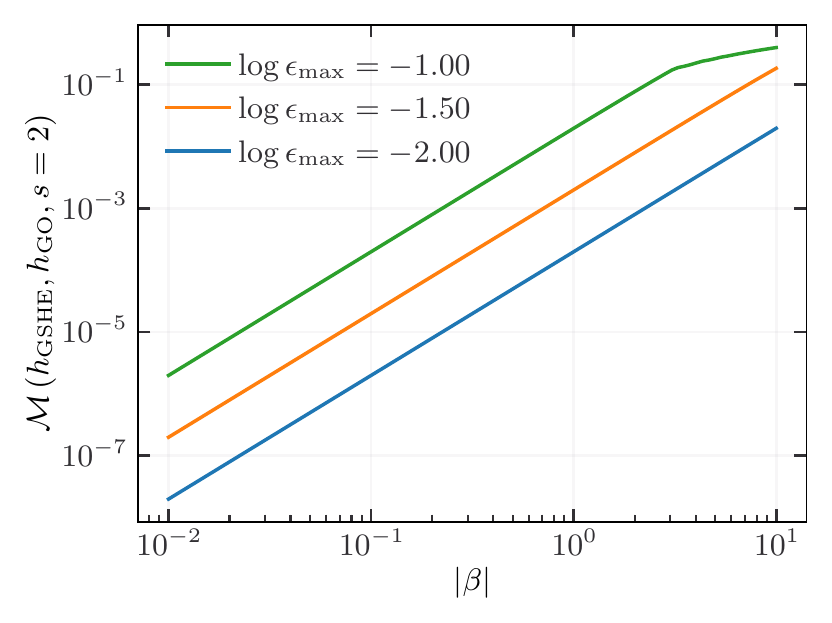}
    \caption{
    The mismatch $\mathcal{M}$, \cref{eq:mismatch}, of the \ac{GSHE}-induced corrections as a function of $|\beta|$ to a single bundle of connecting trajectories for several choices of the maximum perturbation strength $\epsilon_{\rm max}$ and the waveform of~\cref{fig:circ_waveform}.}
    \label{fig:mismatch}
\end{figure}

The terms of order $\epsilon^2$ and higher were neglected in the derivation of the \ac{GSHE}. In this work, we use a maximum value of $\epsilon = 0.1$, at which point we assume that the beyond-linear terms are still negligible. Nevertheless, in~\cref{fig:mismatch} we showed that the effect is significant even when this maximum $\epsilon$ is relaxed. The neglected higher-order contributions are likely to induce wave-like phenomena, such as diffraction, as we depart further from the regime of \ac{GO}. However, the \ac{GSHE} treatment describes the motion of the energy centroid of a wave packet, which is only well defined if $\epsilon \ll 1$. When the wavelength reaches $\epsilon \sim 1$ the \ac{WKB} expansion up to an arbitrary order in $\epsilon$ becomes of little interest, as the perturbation series in $\epsilon$ inevitably breaks down. Therefore, instead of extending the \ac{WKB} analysis to higher orders, it is potentially more instructive to directly solve the linearized gravity perturbation propagation via, e.g., the Teukolsky equation approach~\citep{Teukolsky1973,leite2017absorption,Hayato2021}. This approach was used to study \ac{GW} emission in hierarchical triple systems in Ref.~\cite{Cardoso:2021vjq}. An alternative but no simpler route would be a path integral approach of summing over all possible paths connecting the source and observer, whose extremum would be the classical trajectories considered in this work~~\cite{Feldbrugge:2019fjs}. The upside of the former treatment is its validity up to an arbitrary $\epsilon$. Moreover, it would allow matching the \ac{GSHE} results in an appropriate limit.

%%%%%%%%%%%%%%%%%%%%%%%%%%%%%%%%%%%%%%%%%%%%%%%%%%%%%%%%%%%%%%%%%%%%%%%%%%%%%%%%

\subsection{Source and observer placement}\label{sec:discussion_source}

We assumed that both the observer and the source are static. The assumption of a static, far observer in the Kerr metric is a good approximation if we consider $r_{\rm obs} \rightarrow \infty$, as would be the case for astrophysical observations. Throughout this work, we ensured that our conclusions are independent of the distance of the observer from the \ac{BH}. Additionally, one needs to consider the gravitational and cosmological redshift. We verified that the gravitational redshift due to escaping the strong-field regime of the background \ac{BH} has a negligible dependence on $\epsilon$. It affects both the geodesic and \ac{GSHE} rays equally, and we do not consider it further. The cosmological redshift due to the expansion of the universe is independent of the frequency and, thus, enters as a simple multiplicative factor.

On the other hand, the assumption of a static source may break down, particularly if the source is as close to the \ac{BH} as we have considered above. This depends on the distance traveled by the source while the signal is emitted over the frequency range of a given detector. The former factor depends on the orbital period of the binary around the background \ac{BH}
\begin{equation} \label{eq:orbital_period}
    T_{\rm orb} \approx 138\,s\left(\frac{\mathcal{A}}{10\,R_{\rm s}}\right)^{3/2}\left(\frac{M}{5\times 10^4 M_\odot}\right),
\end{equation}
where $\mathcal{A}$ is the semi-major axis of the orbit. The in-band duration of the signal depends on the \ac{GW} source masses and intrinsic parameters. The typical range of \ac{LVK} in-band source duration are $0.1-100~\mathrm{s}$. The static-source assumption limits the validity of our results to shorter in-band events, including the more massive mergers expected in dynamical formation scenarios and \acp{AGN}. Our framework can be applied to longer events (e.g. lighter sources such as binary neutron star mergers), but only if they orbit a sufficiently massive \ac{BH}, or are located sufficiently far. Source motion also needs to be accounted for if the \ac{GSHE} signature is very sensitive on the source position. This can happen in strongly aligned systems, or for trajectories that undergo a very strong deflection, such as multiple loops around the \ac{BH}.

The static source assumption will be severely violated by stellar mass black holes emitting in the LISA frequency band. These sources have wavelengths several orders of magnitude larger than \ac{LVK} sources. They evolve very slowly in frequency and can be observed over several years~\citep{LISA17, LISA19}, completing multiple orbits around the massive \ac{BH}~\cite{DOrazio:2019fbq}.  A treatment of a moving source would require the composition of the \ac{GSHE} signal across multiple time steps and accounting for the Doppler effects; see Refs.~\citep{Toubiana:2020drf,Sberna:2022qbn}. While the very low frequency ($\sim\mathrm{mHz}$) enhances the \ac{GSHE} corrections, the slow frequency evolution might make a detection challenging. Moreover, at such low frequencies the perturbative expansion in $\epsilon$ may break down, necessitating a treatment in the wave optics regime, unless the background \ac{BH} is sufficiently massive as described in~\cref{eq:epsilon_SI}.

Another potential issue is whether the binary is tidally disrupted by the background \ac{BH}. This can be described by the Hills mechanism~\cite{Hills1988,Addison2019,Suzuki2020} and a significant perturbation occurs when the tidal force induced by the background \ac{BH} is of the same order as the binary's self-gravity. This effect has been estimated in Ref.~\cite{Cardoso:2021vjq} for hierarchical triple systems similar to the ones we are considering. For a binary with an orbital period of $1/f$, tidal effects become important when the binary is at a radius
\begin{equation}
    r_t \lesssim \frac{2 M}{ (M f)^{2/3}} = 2 M \epsilon^{2/3}.
\end{equation}
In this paper, we always consider \acp{GW} with wavelengths smaller than $\epsilon_{\rm max} = 0.1$. Thus, tidal effects can be safely ignored, as they only become significant if the binary is placed at the radius of $r_t \lesssim 0.43 M$, which is below the event horizon of the background \ac{BH}. Thus, binary disruption only affects our results indirectly, by precluding the formation of binaries with $\epsilon\gg 1$, which may later evolve to the range of frequencies probed by \ac{LVK}. Addressing this effect requires detailed considerations on dynamical binary formation and migration beyond the scope of this work.

%%%%%%%%%%%%%%%%%%%%%%%%%%%%%%%%%%%%%%%%%%%%%%%%%%%%%%%%%%%%%%%%%%%%%%%%%%%%%%%%

\subsection{Source emission modeling}\label{sec:disc_anisotropy}

Our analysis relies on a simplified treatment of the \ac{GW} source. Here we comment on the assumptions made: quasi-circular binaries, evolution in vaccuum, and isotropic emission.

We assume a quasi-circular binary system. However, binaries formed in dynamical environments are likely to have non-negligible eccentricity due to three-body interactions~\cite{Samsing:2017xmd,Zevin2021,Samsing2022}.
Even in this case, we expect the eccentricity to be distinguishable from the \ac{GSHE} via its different phase evolution. The phase of eccentric binaries evolves as $\propto f^{-34/9}$~\cite[Eq.~3.13]{Tiwari:2019jtz}, different from the dispersive GSHE, whose phase is modified by $\sim f^{-1}$ ($\Delta t \propto f^{-2}$). Eccentricity can also be distinguished by the presence of higher modes in the signal, which are not induced by the GSHE.
Other environmental effects can be distinguished for similar reasons. Reference~\cite{Toubiana:2020drf} computed environmental corrections for the phase $\propto f^{-13/3}$ (accretion, acceleration) and $f^{-16/3}$ (dynamical friction), distinct from that of the \ac{GSHE}. We also note that these environmental effects are important for low-frequency inspirals, but very suppressed for stellar-mass coalescences.

Any other effect on the emission of \acp{GW} (such as tidal interactions with the central \ac{BH}) can be included in the analysis by updating the unlensed waveform $\tilde h_0(f,s)$ in Eq.~\eqref{eq:GSHEwaveform_frequency}. Even if these effects are partially degenerate, the \ac{GSHE} can be unambiguously discriminated through the existence of a second image with the same underlying waveform but opposite sign of $\beta$.

We have also considered an isotropic \ac{GW} emitter. However, a binary merger is an anisotropic emitter -- similar to an electric dipole -- and the emitted power has a directional dependence (see Ref.~\cite{Gondan:2021fpr} for a treatment of strong-field lensing by Schwarzschild \acp{BH}). There are two effects in which the angular dependence of the source might play a role. First, for a given $\epsilon s$-dependent set of trajectories connecting the source and observer, the initial emission direction must be rotated away from the geodesic emission direction by an angle that is approximately proportional to $\epsilon$. This generally corresponds to an angle of $\sim 1~\mathrm{degrees}$ or lower between the low- and high-frequency components of the signal. This value is well below the sensitivity to the \ac{GW} intrinsic parameters, such as the orbital inclination $\iota$.

The angular structure of the source can cause differences among the images (bundles) formed by the background \ac{BH}. The multiple images may have different relative amplitude, polarization, and merger phase, depending on which angular portion of the binary is projected onto the source for each trajectory. As an example, consider the configuration shown in~\cref{fig:example_trajectory}, in which the two bundles depart in opposite directions from the source. In contrast, each \ac{GSHE} trajectory encompasses an angular deviation proportional to $\epsilon$ relative to the geodesic limit for that bundle. This difference is unrelated to the \ac{GSHE} corrections. However, further studies that quantify the detectability of the \ac{GSHE} will need to explore this effect.

%%%%%%%%%%%%%%%%%%%%%%%%%%%%%%%%%%%%%%%%%%%%%%%%%%%%%%%%%%%%%%%%%%%%%%%%%%%%%%%%

\subsection{Relation to tests of GR}\label{sec:disc_tgr}

If not accounted for, the \ac{GSHE} might be incorrectly interpreted as a deviation from \ac{GR}. In contrast, a detection favoring beyond \ac{GR} physics has to be distinguished from the \ac{GSHE}. Due to its frequency dependence, the \ac{GSHE} mimics three tests of \ac{GR}: a modified dispersion relation, constraints of the post-Newtonian parameters, and consistency of the inspiral, merger, and ringdown phases of the signal. We will focus on the modified dispersion relation, which exactly mimics the \ac{GSHE}-to-geodesic time delay (i.e. $\beta$) if the right-to-left delay is negligible. The connection to the other tests is not straightforward. Hence, we will focus on the modified propagation,~\cref{eq:tgr_prop_0}.

The \ac{GSHE}-induced delay is degenerate with a modified dispersion relation of the form
\begin{equation}\label{eq:tgr_prop_0}
    E^2 = p^2 + c_0,
\end{equation}
in the limit $|c_0/(h f)^2|\ll 1$, where $h$ is Planck's constant. This is a particular case of a generic violation of Lorentz invariance, in which a term proportional to $p^n$ is added ~\cite{LIGOtgrGWTC1,LIGOScientific:2020tif,LIGOScientific:2021sio}.
Our case ($n=0$) is equivalent to a graviton mass $m_g^2=c_0>0$ if the correction has a positive sign. However, the \ac{GSHE} time delay can have either sign depending on the configuration. A modified dispersion causes a frequency-dependent time delay of a \ac{GW} signal
\begin{equation}\label{eq:tgr_delta_t}
    \Delta \tau_{c_0}
    =
    \frac{c_0 D}{(h f)^2} + \mathcal{O}\left(\frac{c_0^2}{h^4 f^4}\right),
\end{equation}
where $D$ is an effective distance to the source that coincides with the standard luminosity distance for low redshift sources~\cite[Eq. 56]{deRham:2016nuf} (see also Ref.~\cite{Will:1997bb}).

Equating~\cref{eq:tgr_delta_t} and~\cref{eq:time-delay-dependence} yields a relation between the \ac{GW} propagation and \ac{GSHE} parameter
\begin{equation}\label{eq:beta_from_lvi_tests}
    \beta
    \approx
    \frac{G}{c^4 h^2} M D c_0
    \approx 0.148 \frac{M}{5\times 10^4 M_\odot}\frac{D}{\text{Gpc}}\frac{c_0}{(10^{-23}\text{eV})^2},
\end{equation}
The \ac{GSHE}-induced delay coefficient can be probed up to a factor $\propto M D$. The effective distance $D$ is related to the source's distance (see Ref.~\cite[Eq. 5]{LIGOtgrGWTC1}), which is constrained by the amplitude of the signal.
In contrast, the mass $M$ of the background \ac{BH} is unknown a-priori. Measuring $M$ would be possible if multiple signals are received, e.g. by measuring their time delay and magnification ratio. For a single signal, it might be possible to constrain $M$ from the orbital acceleration of the binary around the background \ac{BH}, cf.~\cref{eq:orbital_period}. Other means of constraining $M$ may include identifying the source's environment, e.g. via an electromagnetic counterpart, or statistically, e.g. modeling the distribution of mergers around massive \acp{BH}.

The relation in~\cref{eq:beta_from_lvi_tests} allows us to convert \ac{LVK} tests of~\cref{eq:tgr_prop_0} into constraints on $\beta/M$. We use the full posteriors samples from the events analyzed in the third observation run~\cite{LIGOScientific:2020tif,LIGOScientific:2021sio}. The results are shown in~\cref{tab:beta_bounds}, where he show the $90\%$ c.l. (confidence level) for positive and negative values of $c_0$, assuming a fiducial mass of $5\times 10^4M_\odot$. We note that the \ac{LVK} analyses employ a weakly informative prior on $\log(c_0)$, extending many orders of magnitude below the range where the data can probe~\cref{eq:tgr_prop_0}. Therefore, most of the posterior samples lie in a region that is indistinguishable from \ac{GR}, leading to poor sampling of the region where data is informative. An analysis with non-logarithmic priors would lead to more efficient sampling and avoid the need to treat positive and negative values of $\beta/c_0$ separately.

The key difference between a modified dispersion relation of~\cref{eq:tgr_prop_0} and the \ac{GSHE} is that the former is universal: the same coefficient $c_0$ represents a fundamental property of gravity and modifies the waveforms of all \ac{GW} events. On the contrary, the \ac{GSHE} is environmental and the correction is expected to vary between events. Therefore, to constrain $\beta$ from \ac{LVK} bounds on anomalous \ac{GW} propagation, it is necessary to use the bounds on $c_0$ for individual events, rather than the combined value quoted by \ac{LVK}~\cite{LIGOtgrGWTC1,LIGOScientific:2020tif,LIGOScientific:2021sio}. Another consequence is that \ac{GW} propagation tests depend on the source distance, while the \ac{GSHE} does not. Therefore, the $D-c_0$ correlations need to be taken into account when using~\cref{eq:beta_from_lvi_tests} to constrain $\beta$, e.g. using the full posteriors (as in \cref{tab:beta_bounds}).

We note that the birefringent \ac{GSHE} (i.e., polarization-dependent time of arrival due to $\beta_{\rm R-L}$) resembles other beyond-\ac{GR} effects discussed in the literature. Scalar-tensor theories with derivative couplings to curvature~\cite{Bettoni:2016mij} predict that different \ac{GW} (and additional) polarization states travel at different speeds on an inhomogeneous spacetime. This birefringent effect is different from ours in three respects~\cite{Ezquiaga2020}: 1) it involves a difference in the $+/\times$ polarization, rather than R-L (right-to-left), 2) it is independent of frequency, and 3) it depends on the curvature of beyond-\ac{GR} fields, which can be important over astronomical scales, rather than being confined to the vicinity of a compact object. Therefore, the time delay between polarization states associated to these theories is not bounded to any specific scale, and can range from negligible to astronomical, depending on the theory and the lensing configuration. The lack of observation of birefringence in \ac{LVK} data sets stringent bounds on alternative theories \cite{Goyal:2023uvm}. As deviations from \ac{GR} become stronger near a compact object, detecting the \ac{GSHE} imprints for mergers near a massive black hole would set extremely tight bounds on such theories.

Finally, another beyond-\ac{GR} birefringence effect has been studied in Ref.~\cite{Wang2021} as emerging from higher-order corrections to \ac{GR}~\cite{Kostelecky:2003fs,Tasson:2016xib}. Like the \ac{GSHE}, this form of \ac{GW} birefringence involves the circular polarization states and depends on frequency, although it grows with $f$ rather than decaying like the \ac{GSHE}. Moreover, it is again assumed to be a universal property of gravity, rather than an environmental, event-dependent effect. The analysis in Ref.~\cite{Wang2021} showed that all but two \ac{GW} events analyzed were compatible with \ac{GR}. The outliers, GW190521 and GW191109, preferred their form of birefringence over the \ac{GR} prediction. However, one cannot easily interpret this preference as due to the \ac{GSHE}, as a significant $\beta_{\rm R-L}$ is unlikely and an analogue of our, typically larger, \ac{GSHE}-to-geodesic delay due to $\beta$, has not been included in the analysis. Unfortunately, LIGO-Virgo did not quote any results on $c_0$ (\cref{eq:tgr_delta_t}) for that event. Therefore, a more detailed analysis would be required before reaching any conclusions.

\begin{table}
\begin{tabular}{l || l l l l}
\hline
Event & $D_{\rm L}\left[\mathrm{Mpc}\right]$ & $M_{\rm tot}\left[M_\odot\right]$ & $\beta_+$ & $\beta_{-}$ \\
\hline
GW190706 & $5400$ & $190$ & $1.35\times10^{0}$ & $1.2\times10^{-1}$ \\
GW190707 & $780$ & $23$ & $1.5\times10^{-1}$ & $7.5\times10^{0}$ \\
GW190708 & $890$ & $36$ & $1.85\times10^{-1}$ & $8.5\times10^{-1}$ \\
GW190720 & $770$ & $25$ & $3.3\times10^{-1}$ & $3.4\times10^{-1}$ \\
GW190727 & $3000$ & $110$ & $1.75\times10^{-1}$ & $2.1\times10^{-1}$ \\
GW190728 & $920$ & $24$ & $3.9\times10^{-1}$ & $2.4\times10^{-1}$ \\
GW190814 & $300$ & $27$ & $1.3\times10^{-1}$ & $5.0\times10^{-2}$ \\
GW190828 & $2200$ & $80$ & $7.5\times10^{-2}$ & $4.7\times10^{-1}$ \\
GW190910 & $1900$ & $100$ & $4.65\times10^{-2}$ & $3.75\times10^{-1}$ \\
GW190915 & $1700$ & $77$ & $8.5\times10^{-2}$ & $4.3\times10^{-1}$ \\
GW190924 & $580$ & $16$ & $6.0\times10^{0}$ & $1.85\times10^{-1}$ \\
GW191129 & $800$ & $20$ & $6.0\times10^{-1}$ & $4.95\times10^{-1}$ \\
GW191204 & $600$ & $23$ & $1.6\times10^{-1}$ & $5.0\times10^{-2}$ \\
GW191215 & $1900$ & $58$ & $7.0\times10^{-2}$ & $4.05\times10^{-1}$ \\
GW191216 & $360$ & $21$ & $1.5\times10^{0}$ & $6.0\times10^{-2}$ \\
GW191222 & $3100$ & $120$ & $7.0\times10^{-2}$ & $4.4\times10^{-1}$ \\
GW200129 & $870$ & $76$ & $5.5\times10^{-1}$ & $4.6\times10^{-2}$ \\
GW200208 & $2300$ & $92$ & $8.0\times10^{-2}$ & $1.9\times10^{-1}$ \\
GW200219 & $3700$ & $100$ & $7.0\times10^{-2}$ & $4.0\times10^{0}$ \\
GW200224 & $1700$ & $95$ & $8.0\times10^{-2}$ & $1.85\times10^{-1}$ \\
GW200225 & $1100$ & $41$ & $8.5\times10^{-2}$ & $1.7\times10^{1}$ \\
GW200311 & $1100$ & $75$ & $6.0\times10^{-2}$ & $1.85\times10^{-1}$ \\ \hline
\end{tabular}
\caption{90\% c.l. limits on $\beta$ from \ac{LVK} tests of~\cref{eq:tgr_prop_0}, separately for positive and negative values of $c_0$ while assuming background \ac{BH} mass of $5\times 10^4$. We also show the median total mass $M_{\rm tot}$ and luminosity distance $D_{\rm L}$.}
\label{tab:beta_bounds}
\end{table}

%%%%%%%%%%%%%%%%%%%%%%%%%%%%%%%%%%%%%%%%%%%%%%%%%%%%%%%%%%%%%%%%%%%%%%%%%%%%%%%%

\subsection{Detection prospects and applications}\label{sec:disc_detectability}

Throughout this work we considered \ac{GW} sources very close to the background \ac{BH} to illustrate the consequences of the \ac{GSHE} on a waveform. We have focused on the case of a background \ac{BH} in the range of intermediate-mass to massive of $\sim 10^5 M_\odot$. This results in reasonable values of $\epsilon$ that make the \ac{GSHE} detectable for terrestrial observatories. In case of studying the detectability of the \ac{GSHE} with the longer wavelength LISA-like signals, the background \ac{BH} mass would have to be correspondingly increased to achieve similar values of $\epsilon$, such as super-massive \acp{BH}. We expect that there will be a partial degeneracy between the delay proportionality factor $\beta$ and the ratio between the wavelength and the background \ac{BH} mass, as both control the strength of the \ac{GSHE} corrections. Nevertheless, by their definition $\beta$ is independent of frequency, and therefore sufficiently high-quality data should break this degeneracy.

One of the environments to produce promising signals are \acp{AGN}, whose potential is discussed, e.g., in Ref.~\citep{Levin2003,Stone:2016wzz,Samsing2022}. \acp{BH} (and binaries thereof) are expected to migrate radially inward and form binaries~\citep{Bellovary:2015ifg,Secunda:2020mhd,Grishin:2023riv}. This radial migration may bring the \acp{BH} as close as $\sim 6\,R_{\rm s}$ to the background \ac{BH}~\cite{Peng:2021vzr}. Furthermore, migration traps could promote the growth of intermediate-mass \acp{BH} around \acp{AGN}~\cite{McKernan:2012rf}. In addition, a population of intermediate-mass \acp{BH} is expected in globular clusters, although no clear detection is available as of today to constrain their population~\citep{Haberle2021}. We consider \acp{AGN} and globular clusters to be the most likely candidates to host the hierarchical triple systems we consider, although their respective binary \ac{BH} populations also remain poorly constrained~\citep{Ford2021}. Although we have focused mainly on \ac{BH} mergers, neutron star binaries in close proximity to an \ac{AGN} would be ideal to probe the \ac{GSHE}, in addition to nuclear physics \cite{Vijaykumar:2022fst}.

We find there to exist at least two favorable source-observer configurations that result in a strong \ac{GSHE}: aligned and close-by setups. The aligned setup occurs when the source and observer are approximately on opposite sides of the background \ac{BH}. We show in~\cref{fig:shadow_beta} that in this case there exists a ring of initial directions that results in $|\beta| \gtrsim 1$. Because such trajectories converge to a small region opposite the source, they are also magnified, which is represented by the high $|\beta|$ and high magnification cluster of points on~\cref{fig:beta2mu_scatter}.  Additionally, we demonstrate that in this case it is not necessary for the source to be within a few $R_{\rm s}$ of the background \ac{BH}. The sufficient condition is for the trajectories to pass close to the \ac{BH}. In~\cref{fig:upsilon}, we show that the fraction of these initial directions falls approximately as $1 / r_{\rm src}^2$. This is likely to be at least partially balanced by the fact that more mergers may occur from the outer regions of the \ac{AGN} or globular cluster.

The close-by setup occurs for generic source-observer placements, but requires proximity between the source and the background \ac{BH}. Even if the source, \ac{BH} and the observer are not aligned, there is always a strongly deflected connecting bundle that propagates very close to the background \ac{BH} and thus undergoes significant \ac{GSHE} corrections. In~\cref{fig:radial_powerlaw_fit}, we showed that the delay proportionality factor $\beta$ of such bundles tends to a constant, non-negligible value even for large separations between the source and the background \ac{BH}. These trajectories exist in general, but their detectability is limited by demagnification, which is significant for sources far from the background \ac{BH} and/or large deflection angles. Hence, in this setup we expect the \ac{GSHE} to be detectable only for sufficiently close sources, although for most observer locations.

Our scenario predicts the reception of multiple \ac{GW} signals, associated with each of the bundles connecting the source and the observer. The time delay between the signals (bundles) is proportional to the mass of the background \ac{BH}, and together with the relative magnification carries information about the geometry of the system. Furthermore, each image will contain \ac{GSHE} corrections of different strengths.  In the aligned setup, we expect the two magnified images to have only a short time delay between them. The \ac{GSHE} corrections have a sizeable $\beta$, but generally each has an opposite sign, as exemplified in~\cref{fig:configuration_time_delay}. In the close-by setup, we expect to first detect a signal with $\beta\ll 1,\,|\mu|\approx 1$, followed by a demagnified one with a strong \ac{GSHE} (large $|\beta|$, $|\mu|\ll 1$). Unless the source is very close to the background \ac{BH}, the second image will likely appear as a sub-threshold trigger due to exponential demagnification.

The tools developed for the search and identification of strongly lensed \acp{GW}~~\cite{LIGOScientific:2021izm, Dai:2020tpj} can be applied to searches for \ac{GSHE} imprints. A possible approach to find strongly lensed \ac{GW} events is to use the posterior distribution of one image as a prior for the other image, since the two should agree if they describe the same merger~\cite{Janquart2022}. The short time delays between signals involved in our scenario offer two advantages. First, by lowering the chance of an unrelated event being confused as another image~\cite{Caliskan2022wbh} and, secondly, by narrowing down the interval within which to search for sub-threshold triggers carrying a \ac{GSHE} imprint. If the signal contains higher modes, it may be possible to distinguish type II images (saddle points in the lensing potential) from type I/III (local minima/maxima) due to the lensing-induced phase shift~\cite{Dai:2017huk,Ezquiaga2021,Vijaykumar:2022dlp}. This would provide another handle on the lensing setup, as the secondary image (negative parity, lower $\mu$) carries this phase.

\begin{table}
\begin{tabular}{c c c c c}
 Exp. & $M\, [M_\odot]$ & $V_{\mathcal{G}}\, [{\rm Gpc}^3]$ & $z_{\mathcal{G}}$ & $\mathcal{R}^{10 {\rm yr}}_{90}\, [{\rm Gpc}^{-3}{\rm yr}^{-1}]$\\
\hline
LIGO & $10^{4}$ &  $0.10$  & $0.06$ &  $2.41$ \\
  & $10^{5}$ &  $0.01$  & $0.03$ &  $22.92$ \\
  & $10^{6}$ &  $0.25\times 10^{-3}$  & $0.70\times 10^{-2}$ &  $9.04\times 10^{2}$ \\
  & $10^{7}$ &  $0.24\times 10^{-6}$  & $0.67\times 10^{-5}$ &  $9.47\times 10^{5}$ \\
\hline
CE & $10^{4}$ &  $18.58$  & $0.41$ &  $0.01$ \\
  & $10^{5}$ &  $2.40$  & $0.20$ &  $0.10$ \\
  & $10^{6}$ &  $0.37$  & $0.10$ &  $0.62$ \\
  & $10^{7}$ &  $0.28\times 10^{-2}$  & $0.02$ &  $82.36$ \\
\hline
ET & $10^{4}$ &  $30.42$  & $0.50$ &  $0.76\times 10^{-2}$ \\
  & $10^{5}$ &  $3.41$  & $0.22$ &  $0.07$ \\
  & $10^{6}$ &  $0.52$  & $0.12$ &  $0.44$ \\
  & $10^{7}$ &  $0.02$  & $0.04$ &  $13.68$ \\
\hline
    \end{tabular}
    \caption{Effective detection volume and equivalent redshift for different detectors and background \ac{BH} masses. The results assume a $30+30M_\odot$ source at a distance of $r_{\rm src}=5~R_{\rm s}$ from the \ac{BH}, with a detection threshold of $\rho_{\rm thr}=8$. The last column displays the 10-yr 90\% c.l. limits on the merger rate for events with this characteristic, assuming no observation (in units of ${\rm Gpc}^{-3}{\rm yr}^{-1}$).}
    \label{tab:V_GSHE}
\end{table}

The \ac{GSHE} could be used to investigate the environment of \ac{GW} sources. The time delay between signals associated with different bundles can be used to constrain the background \ac{BH} mass $M$, and
$\beta$ can be used to infer the alignment of the source and observer and, potentially, the background \ac{BH} spin. Furthermore, a detection of a nonzero $\beta_{\rm R-L}$ would further indicate a nonzero \ac{BH} spin. In addition, the source's peculiar acceleration may be used to recover information on the mass of the background \ac{BH} if the static-source approximation is broken, cf.~\cref{eq:orbital_period}. If the acceleration can be considered constant, it will impart a $\propto f^2$ correction to the phase, which can be distinguished from the \ac{GSHE}. If the deviation from the static source approximation is dramatic, as expected for LISA stellar-mass sources, much more information about the orbit can be recovered, e.g.~\cite{Sberna:2022qbn}.

\begin{figure*}
    \centering
    \begin{minipage}[t]{0.99\columnwidth}
    \includegraphics[width=\columnwidth]{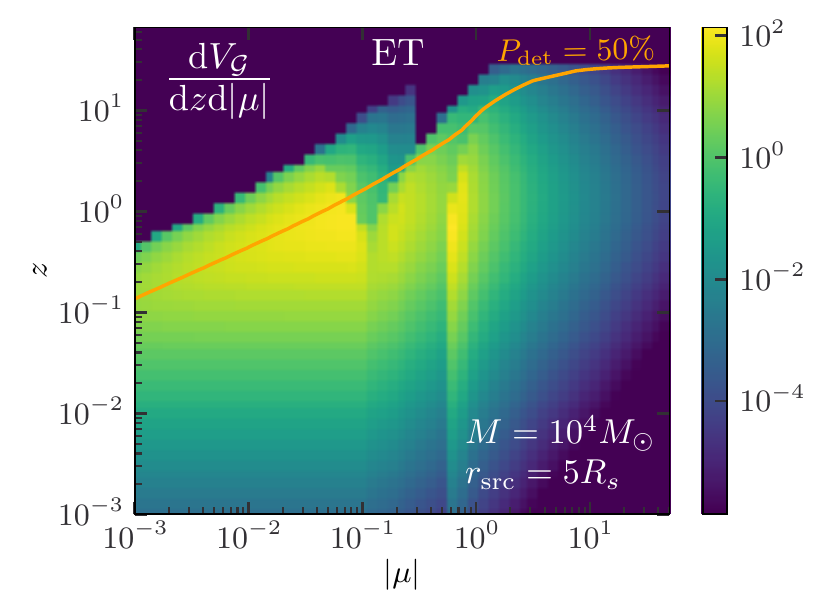}
    \caption{Differential effective volume, \cref{eq:V_GSHE} as a function of the magnification and redshift. The plot applies to a $30+30M_\odot$ binary at $5~R_{\rm s}$ of a $10^4M_\odot$ background \ac{BH}, observed by the Einstein Telescope (see text). The solid line shows the median response distance.}
    \label{fig:V_GSHE}
    \end{minipage}\qquad
    \begin{minipage}[t]{0.99\columnwidth}
    \includegraphics[width=\columnwidth]{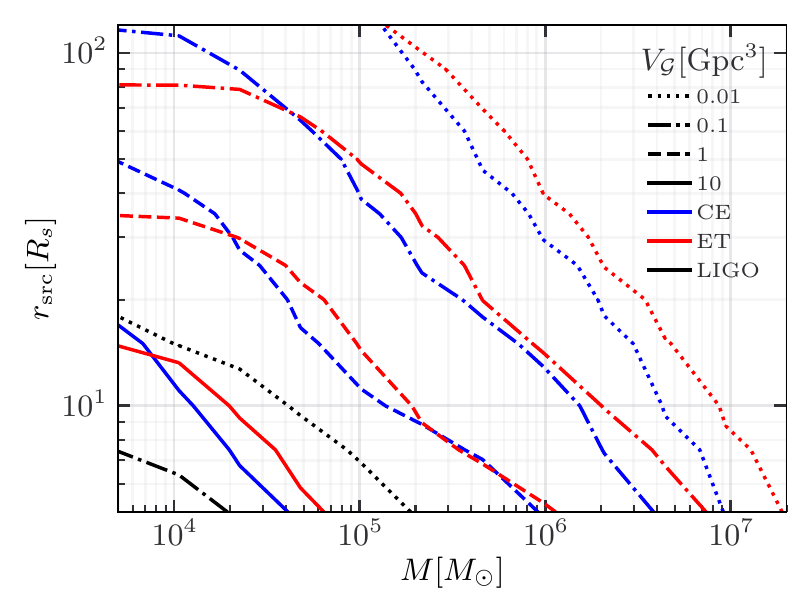}
    \caption{Effective volume \cref{eq:V_GSHE} as a function of the background \ac{BH} mass and the separation of the source. Lines show contours of equal $V_{\mathcal{G}}$ for different detectors.}
    \label{fig:GSHE_volume_all}
    \end{minipage}
\end{figure*}

The capacity to detect \ac{GSHE} corrections in \ac{GW} catalogs remains largely dependent on astrophysical factors. In this exploratory work, we demonstrate that there exist plausible configurations in which the \ac{GSHE} is significant. A detectability study of the \ac{GSHE} would strongly depend on the prior knowledge of the background \ac{BH} population, the merger rates in their environments and their location relative to the background \ac{BH}. We show that the \ac{GSHE}-induced mismatch can reach $\mathcal{M}\sim 10\%$. Under the mismatch and \ac{SNR} criterion that two waveforms are distinguishable if the product $\mathcal{M}\times\text{{SNR}}^2\gtrsim 1$~\citep{Lindblom2008}, we expect \ac{LVK} detectors to find \ac{GSHE} signatures if enough stellar-mass binaries merge in the vicinity of background \acp{BH} of intermediate mass. Recent studies of lensed gamma-ray bursts point towards a population of objects with $M\sim 10^4M_\odot$~\cite{Paynter:2021wmb,Kalantari:2021sqy,Wang:2021ens}, an ideal mass range to observe the \ac{GSHE}.

We now estimate the prospects of \ac{GW} detectors to distinguish the \ac{GSHE} in a signal. To simplify the analysis, we focus on a $30+30 M_\odot$, non-spinning, quasi-circular binary merging at a distance of $r_{\rm src}=5~R_{\rm s}$ from a $10^4 M_\odot$ \ac{BH}. We use the IMRPhenomD waveform model \cite{Husa:2015iqa}, our framework and code for detection probabilities are based on Ref. \cite{Chen:2017wpg}. We consider two setups using the LIGO (O4 curve in Ref.~\citep{KAGRA:2013rdx}), \acl{CE} (\ac{CE};~\citep{Reitze:2019iox}) and \acl{ET} (\ac{ET};~\citep{ET_2020}) noise curves. We assume a single interferometer for simplicity: prospects will improve when considering the \ac{LVK} network, multiple arm combinations in \ac{ET} or a next-generation network of ground detectors \citep{Kalogera:2021bya} thanks to improved \ac{SNR} and sky coverage.

We quantify the observational prospects by defining the effective observable volume as
\begin{equation}\label{eq:V_GSHE}
    V_{\mathcal{G}} = \int \dd z \frac{\dd V_z}{\dd z}(z) \int \dd |\mu| P_{\rm det}  \frac{\dd \Upsilon_{\rm obs}}{\dd |\mu|}.
\end{equation}
Here, $\frac{\dd V_z}{\dd z}(z)$ is the comoving volume element at the source's redshift and $P_{\rm det}(z,|\mu|,\rho_{\rm th})$ is the fraction of signals with \ac{SNR} above a given threshold. The latter depends on the ratio between the detection threshold, $\rho_{\rm th}$, the optimal \ac{SNR} at the source's redshift, $\sqrt{\mu}\rho_{\rm opt}(z)$, and the effect of (de)magnification is shown explicitly. The probability of observable \ac{GSHE}, $\frac{\dd \Upsilon_{\rm obs}}{\dd \mu}(\beta_{\rm min},|\mu|)$, is the derivative of~\cref{eq:upsilon_observer} with respect to $|\mu|$. We further enforce $\frac{\dd \Upsilon_{\rm obs}}{\dd |\mu|}(\beta_{\rm min},|\mu|)\leq 1$, so multiple images contribute at most as one event.  We include all trajectories in our analysis (excluding trajectories with multiple loops has minimal impact on results, which is dominated by strongly deflected trajectories but with with $N_{\rm loop}=0$, cf. \cref{fig:Ups_obs}). The minimum observable value $\beta_{\rm min}$ is determined from the mismatch (Eq. \eqref{eq:mismatch}, \cref{fig:mismatch}) by requiring that $\sqrt{\mathcal{M}(\beta_{\rm min})} > (0.327\rho_{\rm opt})^{-1}$, where the numerical factor relates the optimal \ac{SNR} to the median \ac{SNR}, given $P_{\rm det}$. This threshold, known as the Lindblom criterion \cite{Lindblom:2008cm}, neglects degeneracies between parameters and thus serves as a necessary condition for observability, although it may not be sufficient.

The effective observable volume,~\cref{eq:V_GSHE}, is shown in~\cref{tab:V_GSHE} for different detectors and background \ac{BH} masses. Increasing the \ac{BH} mass severely reduces $V_{\mathcal{G}}$, because only strongly deflected and demagnified trajectories lead to detectable \ac{GSHE}. To facilitate the interpretation, we define an effective redshift so that $V_c(z_{\mathcal{G}}) = V_{\mathcal{G}}$, though it should not be interpreted as a horizon. We can obtain approximate estimates of the number of detections by multiplying $V_{\mathcal{G}}$ by the expected rate $\mathcal{R}$ of events with this characteristics (assuming it is constant) and the observation time $T_{\rm obs}$: $N_{\rm GSHE} = V_{\mathcal{G}}\mathcal{R} T_{\rm obs}$. The probability of detection is described by a Poisson process: in the absence of \ac{GSHE} signatures, the 90\% limit is given by $N_{\rm GSHE}<\ln(0.1)$. \Cref{tab:V_GSHE} shows 90\% c.l. limits on the merger rate of objects at $r_{\rm src}=5~R_{\rm s}$ from the background \ac{BH} of different masses, assuming no \ac{GSHE} detections over an observation period of 10 years.

\Cref{fig:V_GSHE} illustrates the differential effective observable volume, i.e. the integrand of~\cref{eq:V_GSHE} for \ac{CE} with binary masses of $30 + 30~M_\odot$ at a source distance of $5~R_{\rm s}$ from a $10^4~M_\odot$ \ac{BH}. The probability is dominated by strongly deflected but demagnified trajectories, for which \ac{GSHE} distortions are substantial. Highly aligned and magnified trajectories, although less likely, still contribute significantly to detections with $|\mu| > 1$. For \ac{ET} (and similarly \ac{CE}), mildly demagnified trajectories can be observed up to $z\sim 1-10$, at least if the source merges close to the background \ac{BH}.

Figure \ref{fig:GSHE_volume_all} shows $V_{\mathcal{G}}$ for different detectors, as a function of the background \ac{BH} mass and the distance to the source. The scaling of probabilities and magnificatoins with $r_{\rm src}$ employed is described in Sec. \ref{sec:directiondependence}.
The maximum redshift of the detectable region decreases as the mass of the background \ac{BH} increases, since only $\beta\gg 1, |\mu| \ll 1$ trajectories lead to observable signals. However, our estimates are constrained by the resolution of our numerical exploration. A more precise sampling of strongly bent trajectories grazing the lightring will boost the probabilities for $M \gtrsim 10^6~M_\odot$, although detection in those cases is likely to remain difficult even for next-generation ground detectors.

Although the eventual detection of \ac{GSHE} depends on unknown astrophysics, the above results show how prospects will improve dramatically with the next-generation of \ac{GW} detectors. Space detectors sensitive to lower frequencies will provide a great opportunity to probe the \ac{GSHE} in a different regime. LISA, operating in the $\mathrm{mHz}$ window, can detect stellar-mass sources years before merger, including details of their orbit against the background \ac{BH}. The lower frequencies enable our perturbative calculations to yield distinct predictions for binaries orbiting supermassive \acp{BH}, with the caveat that orbital effects need to be included (cf.~\Cref{sec:discussion_source}). The \ac{GSHE} will become most dramatic for a massive background \acp{BH} $\sim 10^6M_\odot$, such as the central \ac{BH} of our galaxy. Large $\epsilon$ may even allow a clear detection of left-to-right birefringence induced by the \ac{GSHE}. However, treating these cases may require a non-perturbative approach (cf.~\Cref{sec:disc_higherordercontribs}). In the future, proposed space-born \ac{GW} detectors will provide new opportunities to search for \ac{GSHE} and wave optics-induced effects on \ac{GW} propagation~\cite{Baker:2019ync,Baibhav:2019rsa,Sedda:2019uro,Sesana:2019vho}.

%%%%%%%%%%%%%%%%%%%%%%%%%%%%%%%%%%%%%%%%%%%%%%%%%%%%%%%%%%%%%%%%%%%%%%%%%%%%%%%%

\section{Conclusion}\label{sec:conclusion}
\acresetall

The \ac{GSHE} describes the propagation of a polarized wave packet of finite frequencies on a background metric in the limit of a small deviation from the \ac{GO} limit. We follow the \ac{GSHE} prescription as presented in Refs.~\cite{GSHE_GW, Harte_2022}. There, the \ac{GSHE} is derived by inserting the \ac{WKB} ansatz into the linearized gravity action and expanding it up to first order in wavelength. The first order contributions include the spin-orbit interaction, resulting in polarization- and frequency-dependent propagation of a wave packet. \ac{GO} is recovered in the limit of infinitesimal wavelength relative to the spacetime characteristic length scale, which in our work is the Schwarzschild radius of the background metric.

The results presented in this work can be framed as a fixed spatial boundary problem. We study the \ac{GSHE}-induced corrections to trajectories connecting a static source and an observer as a function of frequency and polarization. In general, for a fixed source and observer, there exist at least two connecting bundles of trajectories parameterized by $\epsilon  s$, with $\epsilon \equiv 2{\lambda}/{R_{\rm s}}$ and $s=\pm 2$ for \acp{GW}, each of whose infinite frequency limit ($\epsilon \rightarrow 0$) is a geodesic trajectory. There exist additional bundles that loop around the background \ac{BH}. Within each bundle, we compare the time of arrival of the rays as a function of $\epsilon s$ with geodesic propagation.

We find that, regardless of the mutual position of the source and observer or the \ac{BH} spin, the time of arrival delay follows a power law in frequency, with an exponent of $2$ or $3$. The former case corresponds to the dispersive \ac{GSHE}-to-geodesic and the latter to the birefringent right-to-left delay. The information about the relative source-observer position and the polarization is encoded in the power law proportionality constant. The right-to-left delay is suppressed in all but the most extreme configurations, and the time delay of trajectories within a single bundle is, thus, only weakly dependent on the polarization state. Therefore, as an approximation, it can be assumed that the \ac{GSHE} time of arrival is polarization independent and only a function of frequency, i.e. that the time of arrival can be parameterized by $\epsilon$ only instead of $\epsilon s$. Consequently, there is no interference between the right- and left-polarization states, as the difference is negligible for the situations we have studied.

We study the \ac{GSHE}-induced time delay dependence on the relative position of the source and observer, the direction of emission and, lastly, the \ac{BH} spin. We demonstrate that the \ac{GSHE} predicts birefringence effects -- a different time of arrival between right- and left-polarization at a fixed frequency -- only on a spinning Kerr background metric. This is expected from symmetry arguments: the left and right GW polarizations are related by a parity transformation, which would leave a Schwarzschild \ac{BH} invariant, but would flip the spin of a Kerr \ac{BH}.

The \ac{GSHE} corrections to the gravitational waveform manifest as a frequency-dependent phase shift in the inspiral phase of a waveform, the low-frequency components, whose correction is stronger. We compare an example waveform with and without the \ac{GSHE}-induced delay in~\cref{fig:circ_waveform}. We also calculate the \ac{GSHE}-induced waveform mismatch, which can reach $\sim10\%$ in plausible scenarios. Without accounting for the \ac{GSHE} this may be wrongly interpreted as a violation of Lorentz invariance, anomalous \ac{GW} emission or an inconsistency between inspiral-merger-ringdown. Thenceforth, any detection of such an inconsistency must eliminate the \ac{GSHE} before claiming the detection of new physics.

We identify two favorable configurations for detecting the \ac{GSHE}. The first case, an aligned setup, closely mimics the traditional lensing scenario. In it the source and observer are approximately on opposite sides of the background \ac{BH}. In this case, the fraction of initial directions that receive a significant \ac{GSHE} correction falls approximately as $1/r_{\rm src}^2$. The second favorable configuration, a close-by source, follows from relaxing the assumption that the source and the observer are aligned with the background \ac{BH}. In this case, there exist observer-source bundles of trajectories that are strongly deflected by the background \ac{BH} and hence the associated signals have a strong \ac{GSHE} imprint. While these signals are demagnified, they can be observed if the \ac{SNR} of the source is high, it merges sufficiently close to the background \ac{BH}, or both.

These scenarios can be further probed by the existence of multiple lensed signals corresponding to the different \ac{GSHE} bundles. A characteristic signature is that each of the main bundles has opposite signs of the time delay: the first received signal has positive $\beta$, with low frequency components delayed relative to the geodesic. The second signal has negative $\beta$, with low frequency components advanced relative to the geodesic, in addition to a phase shift that might be detected for \ac{GW} sources emitting higher harmonics~\cite{Dai:2017huk,Ezquiaga2021,Vijaykumar:2022dlp}. If current or future \ac{GW} detections reveal \ac{GSHE} imprints, they may be used to constrain the fraction of events near massive and intermediate-mass \acp{BH}, providing further insight into the formation channels of compact binaries.

The \ac{GSHE} is distinct from other wave propagation phenomena, such as diffraction in weak-field lensing \cite{Takahashi:2003ix,Caliskan:2022hbu,Tambalo:2022wlm}. These frequency-dependent modifications of the waveform are associated to lenses at cosmological distances from the source/observer, whose mass is comparable to the GW period. Besides their conceptual differences, both effects can be distinguished in data, as the \ac{GSHE} time delays converge to the geodesic/\ac{GO} limit as $\sim 1/f^2$, more rapidly than the $1/f$ of weak-field diffraction (e.g. interpreting the phase correction in Ref.~\cite[Eq. 11]{Tambalo:2022plm} as a time delay).

The equivalence between the frequency dependence of the \ac{GSHE} and a violation of Lorentz invariance allows us to set limits using existing \ac{LVK} analyses (\cref{tab:beta_bounds}). The 90\% c.l. limits can be as stringent as $|\beta|\lesssim 10^{-2}$, and often differ substantially for positive/negative values of the time delay. Despite potential degeneracies with other waveform parameters, these constraints are in reasonable agreement with expectations based on the mismatch with the geodesic waveform. 

We then analyse detection prospects of current and proposed \ac{GW} detectors on the ground. Next-generation instruments (\acs{ET}, \acs{CE}) have the potential to detected \ac{GSHE} signatures from events near intermediate-mass \acp{BH} ($M\sim 5\times 10^4M_\odot$) if the merger rate within $\sim 25 R_{\rm s}$ is $\mathcal{O}(1)~{\rm Gpc}^{-3}{\rm yr}^{-1}$. These estimates are conservative, as they consider a single interferometer and are limited by the resolution of our numerical studies for trajectories grazing the background \ac{BH}, which dominate the probability. The sensitivity drops sharply for larger masses and separations; however, upcoming instruments in space such as LISA~\cite{LISA17,LISA19}, TianQin and Taiji~\cite{Gong:2021gvw} in the 2030s and proposals in the decihertz~\cite{Sedda:2019uro}, milihertz~\cite{Baibhav:2019rsa} and microhetz~\cite{Sesana:2019vho} bands offer the best prospect for observing the \ac{GSHE}. Addressing the full phenomenology of the \ac{GSHE} and its detectability by next-generation detectors will require extending our formalism for non-static sources and beyond the \ac{GO} expansion.

We conclude that there exists potential to unambiguously detect the \ac{GSHE}. This hints at an optimistic future for studying the \acl{GW} propagation in strong gravitational fields, novel tests of \acl{GR} and decoding imprints of the merger environment (e.g. the spin of the lens \ac{BH} if the birefringent \ac{GSHE} is observable) directly from individual waveforms.

%%%%%%%%%%%%%%%%%%%%%%%%%%%%%%%%%%%%%%%%%%%%%%%%%%%%%%%%%%%%%%%%%%%%%%%%%%%%%%%%

\section*{Acknowledgements}

We thank Lars Andersson, Pedro Cunha, Dan D'Orazio, Bence Kocsis, Johan Samsing, Laura Sberna, and Jochen Weller for input and discussions, as well as the anonymous referees for constructive criticism.
RS acknowledges financial support from STFC Grant No. ST/X508664/1 and the Deutscher Akademischer Austauschdienst (DAAD) Study Scholarship.

This research has made use of data, software and/or web tools obtained from the Gravitational Wave Open Science Center (\url{https://www.gw-openscience.org/}), a service of the LIGO Laboratory, the LIGO Scientific Collaboration and the Virgo Collaboration. LIGO is funded by the U.S. National Science Foundation. Virgo is funded, through the European Gravitational Observatory (EGO), by the French Centre National de Recherche Scientifique (CNRS), the Italian Istituto Nazionale della Fisica Nucleare (INFN) and the Dutch Nikhef, with contributions by institutions from Belgium, Germany, Greece, Hungary, Ireland, Japan, Monaco, Poland, Portugal, Spain.

\appendix
\renewcommand{\theequation}{\thesection.\arabic{equation}}

%%%%%%%%%%%%%%%%%%%%%%%%%%%%%%%%%%%%%%%%%%%%%%%%%%%%%%%%%%%%%%%%%%%%%%%%%%%%%%%%

\section{Proper time and orthonormal tetrad}

%%%%%%%%%%%%%%%%%%%%%%%%%%%%%%%%%%%%%%%%%%%%%%%%%%%%%%%%%%%%%%%%%%%%%%%%%%%%%%%%

\subsection{Observer proper time}\label{sec:observer_proper_time}

\begin{figure*}
    \centering
    \includegraphics[width=\textwidth]{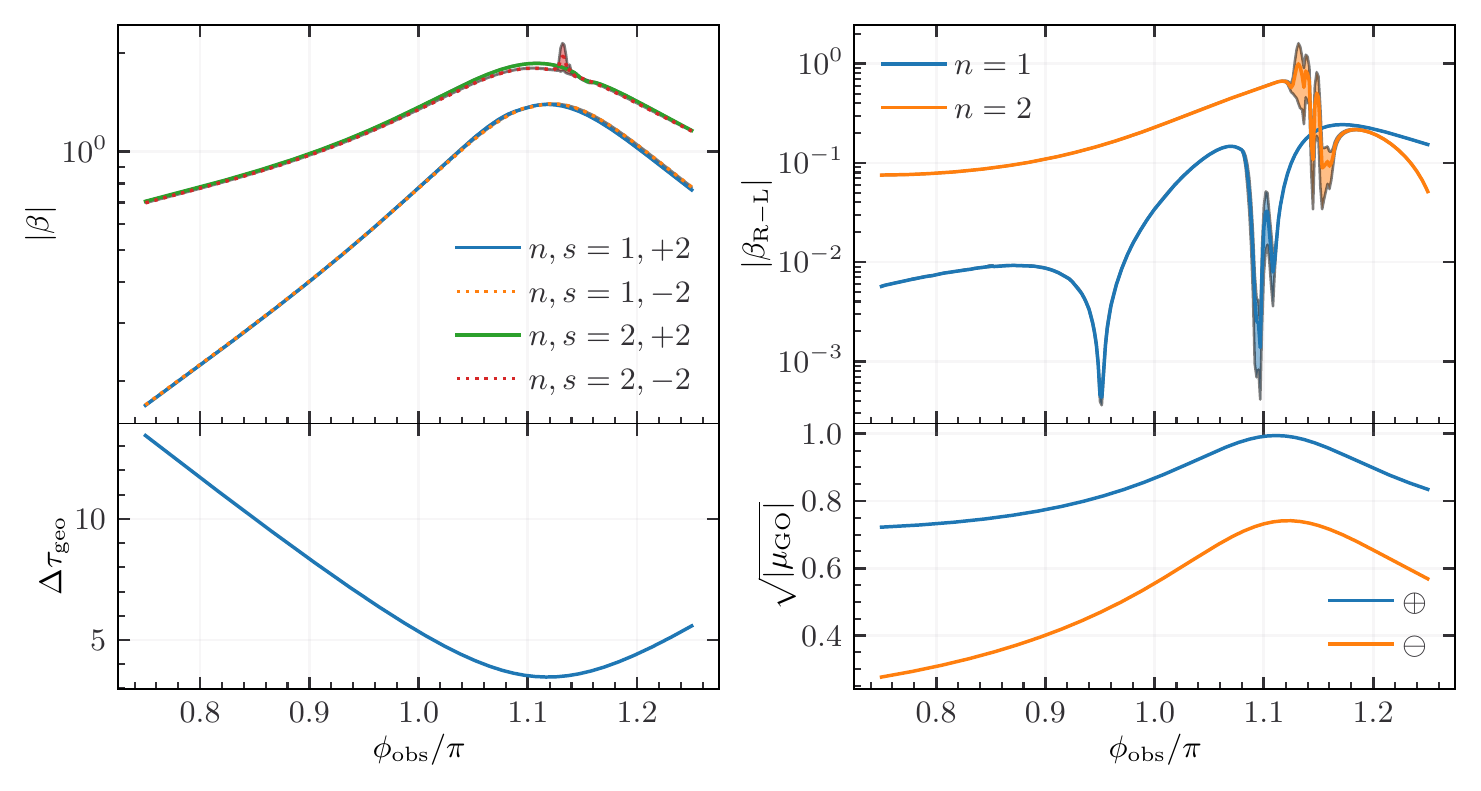}
    \caption{Time delay parametrization upon varying the observer azimuthal angle $\phi_{\rm obs}$. The \emph{top} row shows $\beta$ and $\beta_{\rm R-L}$. The \emph{bottom} row shows $\Delta \tau_{\rm geo}$ and $\mu_{\rm GO}$. The source is otherwise at $(2\,R_{\rm s}, \pi/2, 0)$, observer at $(50\,R_{\rm s}, 0.4\pi, \phi_{\rm obs})$ and $a=0.99$. We still have that $\alpha \approx 2$ and $\alpha_{\rm R-L} \approx 3$.}
    \label{fig:phiobs_powerlaw}
\end{figure*}

\begin{figure*}
    \centering
    \includegraphics[width=\textwidth]{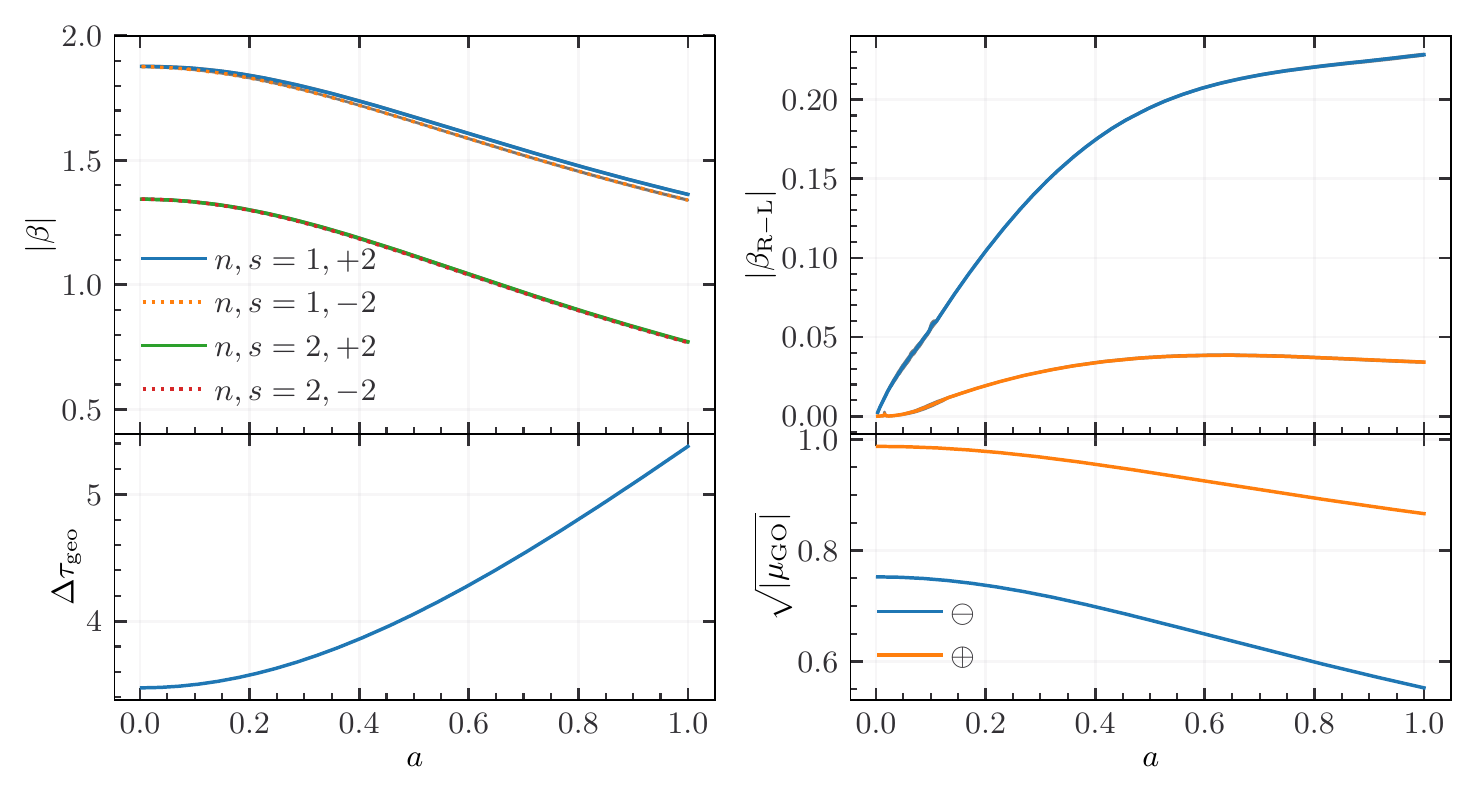}
    \caption{Time delay parametrization upon varying the \ac{BH} spin $a$. The \emph{top} row shows $\beta$ and $\beta_{\rm R-L}$. The \emph{bottom} row shows $\Delta \tau_{\rm geo}$ and $\mu_{\rm GO}$. The source is at $(2\,R_{\rm s}, \pi/2, 0)$ and the observer at $(50\,R_{\rm s}, 0.4\pi, \pi)$. We again have that $\alpha \approx 2$ and $\alpha_{\rm R-L} \approx 3$. In the Schwarzschild metric the \ac{GSHE}-to-geodesic delay is maximized, while the right-to-left delay is zero.}
    \label{fig:BHspin_beta}
\end{figure*}

We assume the far static observer to follow a worldline $\gamma_{\rm obs}(\tau)$ parameterized in the Boyer-Lindquist coordinate system of a Kerr metric as
\begin{equation}\label{eq:observer_worldline}
    (\gamma_{\rm obs})^\mu (\tau)
    =
    \left( t_{\rm obs}(\tau), r_{\rm obs}, \theta_{\rm obs}, \phi_{\rm obs} \right),
\end{equation}
where the spatial coordinates $\bm{x}_{\rm obs} = (r_{\rm obs}, \theta_{\rm obs}, \phi_{\rm obs})$ are constant. Therefore, the $4$-velocity of this observer is
\begin{equation}\label{eq:observer_4velocity}
    \dv{(\gamma_{\rm obs})^\mu}{\tau}
    =
    \left( \dv{t_{\rm obs}(\tau)}{\tau}, 0, 0, 0 \right),
\end{equation}
and to ensure that $\gamma_{\rm obs}(\tau)$ is parameterized in terms of observer's proper time $\tau$ we impose that
\begin{equation}
    \tensor{g}{_\mu _\nu} \dv{(\gamma_{\rm obs})^\mu}{\tau} \dv{(\gamma_{\rm obs})^\mu}{\tau}
    =
    \left.\tensor{g}{_0 _0}\right|_{\bm{x}_{\rm obs}} \left(\dv{t_{\rm obs}}{\tau}\right)^2
    =
    -1.
\end{equation}
From the above equation, together with the assumption that $\dd \gamma_{\rm obs} / \dd \tau$ is future-directed with respect to the Killing vector field $\partial_t$, we obtain
\begin{equation}\label{eq:observer_time}
    \tau
    =
    t_{\rm obs} \sqrt{-\left.\tensor{g}{_0 _0}\right|_{\bm{x}_{\rm obs}}},
\end{equation}
which, up to a constant addition factor, relates the coordinate time to the observer's proper time.

%%%%%%%%%%%%%%%%%%%%%%%%%%%%%%%%%%%%%%%%%%%%%%%%%%%%%%%%%%%%%%%%%%%%%%%%%%%%%%%%

\subsection{Alignment of an arbitrary tetrad}\label{sec:tetrad}

We consider another orthonormal tetrad $\tilde{e}_a$ related to $e_a$ by spacetime-dependent boosts, with boost velocity $\bm{v} = (v_1, v_2, v_3)$. The boosted orthonormal tetrad $\tilde{e}_a$ can be defined as in Ref.~\cite[Eq. 9]{Grenzebach2015}
\begin{subequations} \label{eq:general_boosted_tetrad}
\begin{align}
    \tilde{e}_0 &= \frac{e_0 + v_1 e_1 + v_2 e_2 + v_3 e_3}{\sqrt{1 - v^2}},\\
    \tilde{e}_1 &= \frac{(1 - {v_2}^2) e_1 + v_1 (v_2 e_2 + e_0)}{\sqrt{1-{v_2}^2}\sqrt{1-{v_1}^2-{v_2}^2}},\\
    \tilde{e}_2 &= \frac{e_2 + v_2 e_0}{\sqrt{1 - {v_2}^2}},\\
    \tilde{e}_3 &= \frac{(1 - {v_1}^2 - {v_2}^2)e_3 + v_3 (v_1 e_1 + v_2 e_2 + e_0)}{\sqrt{1 - {v_1}^2 - {v_2}^2}\sqrt{1 - v^2}},
\end{align}
\end{subequations}
where $v^2 = {v_1}^2 + {v_2}^2 + {v_3}^2 < 1$. In our case, we consider the original orthonormal tetrad $e_a$ to be that of the Kerr metric defined in~\cref{eq:kerr_orthonormal_tetrad}.

As discussed in~\Cref{sec:observer_proper_time}, a static observer follows a worldline $\gamma_{\rm obs}(\tau)$. We wish to align $e_0$ with $\tilde{e}_0$ so that
\begin{equation}
    (\tilde{e}_0)^\mu
    = \dv{(\gamma_{\rm obs})^\mu}{\tau}
    =1/\sqrt{-\left.\tensor{g}{_0 _0}\right|_{\bm{x}_{\rm obs}}} \tensor{\delta}{^\mu _0}.
\end{equation}
Therefore, we will need a boost with $\mathbf{v} = (0, 0, v_3)$, where
\begin{equation}
    v_3 = - \frac{a \sin{\theta_{\rm obs}}}{ \sqrt{\Delta(r_{\rm obs})}} e^{-(r - r_{\rm obs})^2},
\end{equation}
where the exponential ensures a smooth alignment from $e_a$ far from the observer to $\tilde{e}_a$ at the observer's position. A similar boost can be performed at the source's location, which will be valid as long as the two exponentials have no significant overlap.

%%%%%%%%%%%%%%%%%%%%%%%%%%%%%%%%%%%%%%%%%%%%%%%%%%%%%%%%%%%%%%%%%%%%%%%%%%%%%%%%

\section{Additional time delay scaling}\label{sec:more_timedelay}

We now continue with the discussion from~\Cref{sec:numerical_time_delay} of the \ac{GSHE}-induced time delay with respect to the geodesic arrival as a function of the azimuthal separation and the Kerr \ac{BH} spin.

%%%%%%%%%%%%%%%%%%%%%%%%%%%%%%%%%%%%%%%%%%%%%%%%%%%%%%%%%%%%%%%%%%%%%%%%%%%%%%%%

\subsection{Dependence on the azimuthal separation}

In~\cref{fig:phiobs_powerlaw}, we vary the azimuthal angle of the observer $\phi_{\rm obs}$. The source is at $(2\,R_{\rm s}, \pi/2, 0)$, the observer is at $(50\,R_{\rm s}, 0.4\pi, \phi_{\rm obs})$ and $a=0.99$. Because of the nonzero \ac{BH} spin, the setup is not symmetric around $\phi_{\rm obs} = \pi$ and instead we find that trajectories moving against the direction of the \ac{BH} spin receive stronger \ac{GSHE} corrections. There exist symmetric source-observer configurations in which the right-to-left delay appears to vanish, although at present we do not investigate their origin further. When $\phi_{\rm obs} \approx = 1.1\pi$ the \ac{GSHE} corrections and geodesic magnification are maximized. Changing the sign of the \ac{BH} spin, this point moves as expected to $0.9\pi$.

%%%%%%%%%%%%%%%%%%%%%%%%%%%%%%%%%%%%%%%%%%%%%%%%%%%%%%%%%%%%%%%%%%%%%%%%%%%%%%%%

\subsection{Dependence on the BH spin}

In~\cref{fig:BHspin_beta}, we plot how the time delay parameters of directly connecting bundles -- $\beta$, $\beta_{\rm R-L}$, $\Delta \tau_{\rm geo}$ and $\mu_{\rm GO}$ -- depend on the \ac{BH} spin $a$ while keeping the source and observer fixed. The source is placed at $(2\,R_{\rm s}, \pi/2, 0)$ and the observer at $(50\,R_{\rm s}, 0.4\pi, \pi)$. We again note that in all cases $\alpha \approx 2$ and $\alpha_{\rm R-L} \approx 3$. The \ac{GSHE}-to-geodesic delay is maximized when the Kerr metric approaches the Schwarzschild limit, while the right-to-left delay vanishes in the Schwarzschild metric. We attribute the Schwarzschild maximum of the \ac{GSHE}-to-geodesic delay to the fact that the Kerr \ac{BH} horizon grows with decreasing spin, and, therefore, the trajectories pass closer to it. There is no \ac{GSHE} birefringence if the black hole is not spinning because of the reflection symmetry of the Schwarzschild metric. Lastly, we verify that this behavior is not a consequence of a particular source-observer configuration and qualitatively holds in general.

\begin{figure*}
    \centering
    \begin{minipage}[t]{0.99\textwidth}
    \includegraphics[width=\textwidth]{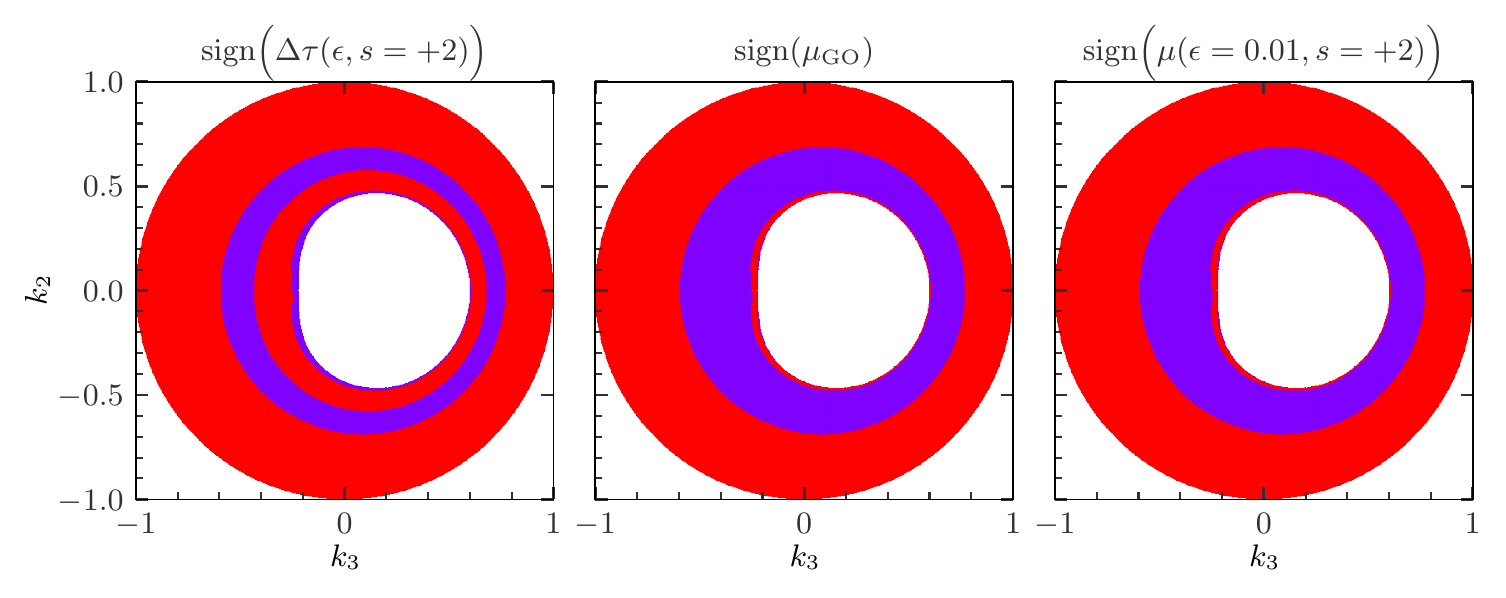}
    \caption{Comparison of the \ac{GSHE}-to-geodesic delay sign (\emph{left} panel), the geodesic parity (\emph{middle} panel) and the \ac{GSHE} parity (\emph{right} panel). Red and violet denote $+1$ and $-1$, respectively. This is plotted as a function of the initial momenta for a source at $(5\,R_{\rm s}, \pi/2, 0)$ and $a=0.99$, corresponding to~\cref{fig:shadow_beta}. The \ac{GSHE}-to-geodesic delay sign is in agreement with the image parity everywhere except in the central red region of the left panel. The \ac{GSHE} parity approximately agrees with the geodesic parity.}
    \label{fig:parity_comp}
    \end{minipage}\qquad 
    
    \vspace{0.5cm}
    
    \begin{minipage}[t]{0.99\columnwidth}
    \includegraphics[width=\columnwidth]{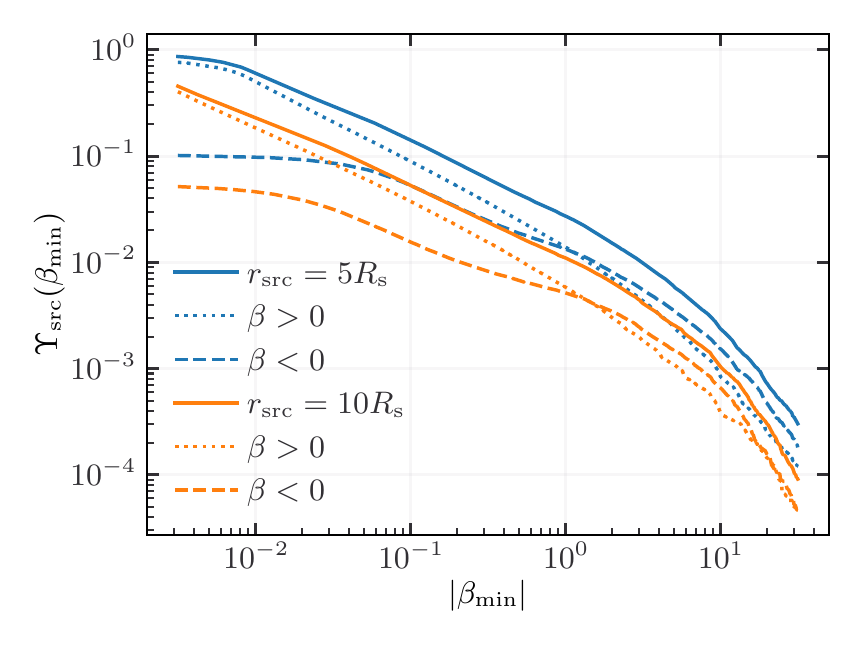}
    \caption{Effect of the sign of $\beta$ on the source's probability at different source radii.}
    \label{fig:ups_source_beta}
    \end{minipage}\qquad
    \begin{minipage}[t]{0.99\columnwidth}
    \includegraphics[width=\columnwidth]{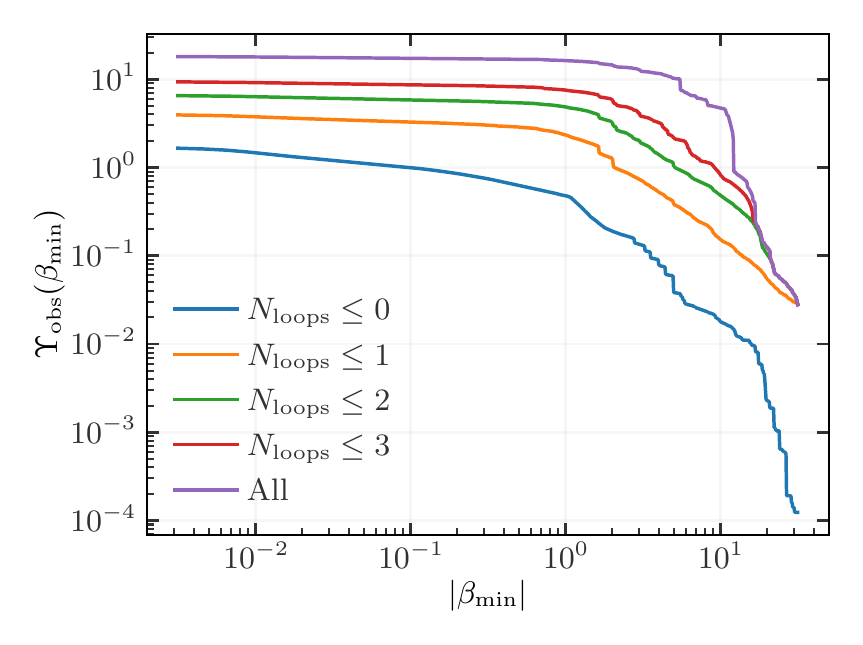}
    \caption{Effect of the number of loops on the observer's GSHE probability, for any value of $|\mu|$.}
    \label{fig:ups_obs_Nloop}
    \end{minipage}
\end{figure*}

%%%%%%%%%%%%%%%%%%%%%%%%%%%%%%%%%%%%%%%%%%%%%%%%%%%%%%%%%%%%%%%%%%%%%%%%%%%%%%%%

\section{Relation between the image parity and the GSHE-to-geodesic delay sign}\label{sec:parity_sign}

We investigate the relationship between signal parity and the sign of the \ac{GSHE}-to-geodesic delay $\Delta \tau(\epsilon, s)$ in the settings of~\cref{fig:shadow_beta}, where we previously calculated $\beta$ as a function of the direction of emission (cf.~\Cref{sec:directiondependence}). We show this in~\cref{fig:parity_comp} for a right-polarized wave ($s=2$), where the red and violet regions correspond to $+1$ and $-1$, respectively. First, in the left panel, the negative time delay corresponds to a well-defined region whose outside boundary is the Kerr equivalent of the Einstein ring, where the determinant of the trajectory mapping approaches zero, or, equivalently, the magnification tends to infinity. This boundary also delineates the middle panel which shows the geodesic parity. However, unlike in the left panel, the negative geodesic parity region extends almost to the \ac{BH} shadow boundary, where the parity starts to oscillate as the solutions begin to completely loop around the \ac{BH}. Thus, outside of this, it is only within the central red region of the left panel where the \ac{GSHE}-to-geodesic delay and image parity signs disagree. For completeness, we also include the magnification at a finite value of $\epsilon$ in the right panel, although it is nearly indistinguishable from the geodesic magnification. Due to the weak dependence on $\epsilon$, we do not expect this picture to change qualitatively for the left polarization. Our results are in agreement with the theory of standard lensing in the weak limit, where trajectories outside the Einstein ring have positive parity and negative parity inside it~\citep{gravitational_lenses_book,dodelson_2017}.

%%%%%%%%%%%%%%%%%%%%%%%%%%%%%%%%%%%%%%%%%%%%%%%%%%%%%%%%%%%%%%%%%%%%%%%%%%%%%%%%

\section{Sign dependence of $\beta$ and  $N_{\rm loop}$} \label{sec:upsilon_details}

In this section, we explore some additional details of the \ac{GSHE} detection likelihood. \cref{fig:ups_source_beta} shows the dependence of the source's probabilities on the sign of the time delay. Negative \ac{GSHE} time delays ($\beta < 0$) are less likely when the effect is small, but larger when $\beta_{\min} \gtrsim 1$. As the sign of $\beta$ can be distinguished by observation, this information can be included when computing probabilities.

\cref{fig:ups_obs_Nloop} depicts the effect of limiting the number of loops on the observer's probability, without any limit on $|\mu|$. This plot differs from~\cref{fig:Ups_obs} where no cut on the magnification is employed: trajectories with multiple loops are strongly demagnified ($|\mu|\ll 10^{-3}$), and therefore ``spread'' widely over all possible observers, leading to $\Upsilon_{\rm obs} \sim \mathcal{O}(10)$, even for $|\beta| \gtrsim 1$. Notably, each number of loops contains two families of trajectories (positive and negative parity). As $\beta_{\min} \to 0$, the $N_{\rm loop}=0$ case approaches the expected value $3/2$, which corresponds to two families of trajectories, with one of them (with weaker deflection) covering only half of the sphere. However, a similar calculation in the multi-loop case is limited by the spatial resolution of our simulations.

%%%%%%%%%%%%%%%%%%%%%%%%%%%%%%%%%%%%%%%%%%%%%%%%%%%%%%%%%%%%%%%%%%%%%%%%%%%%%%%%

\bibliography{references}

\end{document}